\PassOptionsToPackage{svgnames}{xcolor}
\documentclass[11pt]{article}

\usepackage{amsmath,amssymb,amsfonts,amsthm,setspace,tikz,natbib,dsfont, xcolor, array,colortbl,mathrsfs,pifont,wasysym}
\usetikzlibrary{calc}
\usetikzlibrary{positioning}
\usepackage{centernot}
\usepackage{xspace}
\usepackage{mathtools}

\usepackage{thmtools,thm-restate}
\usepackage{enumerate, enumitem}

\usepackage[makeroom]{cancel}

\usepackage[letterpaper]{geometry}
\definecolor{lam1}{HTML}{900C3F}
\definecolor{lam2}{HTML}{364D73}

\usepackage[colorlinks=true,urlcolor=lam2,linkcolor=lam2,citecolor=lam2]{hyperref}

\newtheoremstyle{mytheoremstyle} % name
    {\topsep}                    % Space above
    {\topsep}                    % Space below
    {\itshape}                   % Body font
    {}                           % Indent amount
    {\sc}                   % Theorem head font
    {.}                          % Punctuation after theorem head
    {.5em}                       % Space after theorem head
    {}  % Theorem head spec (can be left empty, meaning ‘normal’)

\theoremstyle{mytheoremstyle}

\newtheorem{theorem}{Theorem}
\newtheorem{lemma}{Lemma}
\newtheorem{proposition}[theorem]{Proposition}
\newtheorem{corollary}{Corollary}

\newtheoremstyle{scfont} % name
    {\topsep}                    % Space above
    {\topsep}                    % Space below
    {}                   % Body font
    {}                           % Indent amount
    {\scshape}                   % Theorem head font
    {}                          % Punctuation after theorem head
    {.5em}                       % Space after theorem head
    {\textbf{Axiom \thmnumber{#2}\thmname{#1}}\thmnote{---{\textsc{#3}.}}}

\theoremstyle{scfont}

\theoremstyle{remark}   
\newtheorem{remark}{Remark}

\newenvironment{tproof}[1]{\noindent \emph{Proof of Theorem \ref{#1}. \phantomsection\label{pf:#1}}}{\hfill$\blacksquare$ \\}
\newenvironment{pproof}[1]{\noindent \emph{Proof of Proposition \ref{#1}. \phantomsection\label{pf:#1}}}{\hfill$\blacksquare$ \\}

\newenvironment{lproof}[1]{\noindent \emph{Proof of Lemma \phantomsection\ref{#1}.}}{\hfill$\blacksquare$ \\}

\newcounter{ax}
\newenvironment{ax}[3]{
\medskip
\addtocounter{ax}{1}
\noindent \textbf{#1\theax}---\sclabel{#2}{#3}.
}{
\medskip
}

\newenvironment{ax'}[3]{
\medskip
\addtocounter{ax}{1}
\noindent \textbf{#1\theax'}---\sclabel{#2}{#3}.
}{
\medskip
}

\newcounter{ex}
\newenvironment{ex}[1]{
\medskip
\refstepcounter{ex}
\noindent \textbf{Example \theex.} (#1)
}{
\qed
\medskip
}

\usepackage{titlesec}

\titleformat*{\section}{\large\scshape\centering}
\titleformat*{\subsection}{\scshape\centering}
\titleformat*{\subsubsection}{\itshape}
\titleformat*{\paragraph}{\large\bfseries\centering}
\titleformat*{\subparagraph}{\large\bfseries\centering}

\titlespacing*{\section}{0pt}{5.5ex plus 1ex minus .2ex}{3.5ex plus .2ex}
\renewcommand{\emptyset}{\varnothing}
\renewcommand{\phi}{\varphi}

\newcommand{\commentout}[1]{}

\renewcommand{\implies}{\, {\Rightarrow}\, }
\renewcommand{\iff}{\, {\Leftrightarrow}\, }

\def \R{\mathbb{R}}

\def \W{\Omega}
\def \w{\omega}
\def \n{\textsc{not}}
\def \P{\pi}

\def \l{\!\land\!}
\renewcommand{\d}[1][{\implies}]{[\![#1]\!]}

\def \F{\mathscr{F}}

\def \B{\mathcal{B}}

\def \<{\langle}
\def \>{\rangle}

\def \s{\succcurlyeq}

\def \1{\mathds{1}}

\makeatletter
\newcommand{\shorteq}{%
  \settowidth{\@tempdima}{..}% Width of hyphen
  \resizebox{\@tempdima}{\height}{=}%
}
\makeatother

\DeclareMathOperator \im{\shorteq\!\!\s}

\makeatletter
\newcommand{\mylabel}[3]{\def\@currentlabel{#2}\phantomsection\textsc{#3} (\texttt{#2})\label{#1}}
\newcommand{\sclabel}[2]{\def\@currentlabel{\theax}\phantomsection\lowercase{\textsc{#2}} (\textsc{#1})\label{ax:#1}}
\makeatother

\renewcommand{\r}[1]{\hyperref[#1]{\textup{(\texttt{\ref{#1}})}}}
\newcommand{\rax}[1]{\hyperref[ax:#1]{\textup{(\textsc{#1})}}}

\newcounter{inline}

\newcommand{\hide}[1]
{
}

\title{Hypothetical Expected Utility}

\author{
Evan Piermont\footnote{Royal Holloway--University of London, Department of Economics, United Kingdom; \href{mailto:evan.piermont@rhul.ac.uk}{\tt evan.piermont@rhul.ac.uk}.}
} 

\begin{document}

\maketitle

\begin{abstract}
\footnotesize
This paper provides a model to analyze and identify a decision maker's hypothetical reasoning. Using this model, I show that a DM's propensity to engage in hypothetical thinking is captured exactly by her ability to recognize implications (i.e., to identify that one hypothesis implies another) and that this later relation is encoded by a DM's observable behavior. Thus, this characterization both provides a concrete definition of (flawed) hypothetical reasoning and, importantly, yields a methodology to identify these judgments from standard economic data. 

\vspace{0.2cm}

\end{abstract}

%%%%%%%%%%%%%%%%%%%%%%%%%%%%%%%%%%%%%%%%%%%%%%%%%%%%
%%%%%%%%%%%%%%%%%%%%%%%%%%%%%%%%%%%%%%%%%%%%%%%%%%%%
%%%%%%%%%%%%%%%%%%%%%%%%%%%%%%%%%%%%%%%%%%%%%%%%%%%%
%%%%%%%%%%%%%%%%%%%%%%%%%%%%%%%%%%%%%%%%%%%%%%%%%%%%
%%%%%%%%%%%%%%%%%%%%%%%%%%%%%%%%%%%%%%%%%%%%%%%%%%%%
%%%%%%%%%%%%%%%%%%%%%%%%%%%%%%%%%%%%%%%%%%%%%%%%%%%%
%%%%%%%%%%%%%%%%%%%%%%%%%%%%%%%%%%%%%%%%%%%%%%%%%%%%
%%%%%%%%%%%%%%%%%%%%%%%%%%%%%%%%%%%%%%%%%%%%%%%%%%%%
%%%%%%%%%%%%%%%%%%%%%%%%%%%%%%%%%%%%%%%%%%%%%%%%%%%%
%%%%%%%%%%%%%%%%%%%%%%%%%%%%%%%%%%%%%%%%%%%%%%%%%%%%
%%%%%%%%%%%%%%%%%%%%%%%%%%%%%%%%%%%%%%%%%%%%%%%%%%%%
%%%%%%%%%%%%%%%%%%%%%%%%%%%%%%%%%%%%%%%%%%%%%%%%%%%%
%%%%%%%%%%%%%%%%%%%%%%%%%%%%%%%%%%%%%%%%%%%%%%%%%%%%
%%%%%%%%%%%%%%%%%%%%%%%%%%%%%%%%%%%%%%%%%%%%%%%%%%%%
%%%%%%%%%%%%%%%%%%%%%%%%%%%%%%%%%%%%%%%%%%%%%%%%%%%%
%%%%%%%%%%%%%%%%%%%%%%%%%%%%%%%%%%%%%%%%%%%%%%%%%%%%

\section{Introduction}
\label{section:motivation}

In choosing between complex alternatives, a decision maker (DM) must often think hypothetically. That is, she must consider the outcome of her choices  contingent on different possible eventualities. Evidence from experimental economics and psychology, reinforcing conventional wisdom, shows that DMs often make costly errors as a result of failures in hypothetical reasoning, for example regarding auctions and procurement  \citep{thaler1988anomalies, li2017obviously, martin2020contingencies}, voting \citep{feddersen2004rational,esponda2014hypothetical}, signaling and disclosure \citep{jin2015no}, information acquisition and belief updating \citep{enke2019correlation, enke2020you}, choice under risk \citep{agranov2020non} choice under uncertainty \citep{martinez2019failures,esponda2019contingent}, the construction of subjective likelihoods  \citep{tversky1983extensional, tversky1994support} generalized strategic inference \citep{eyster2005cursed, esponda2008behavioral}, etc.
%Indeed, distractible, computationally limited, psychologically biased, and otherwise imperfect decision makers will exercise choices in line with their flawed perception of the decision problem at hand.
%In contrast to many biases in decision making (uncertainty aversion, improper Bayesian updating, reference dependence, etc) there are few, if any, simple yet general models of flawed hypothetical reasoning. 
As such, the soundness and precision of our economic models rest acutely on our modeling the strategic consideration of agents not only as they actually are, but also as they are perceived by the agents themselves: making solid predictions in a world with boundedly rational strategic agents requires a testable and identifiable model of hypothetical reasoning. Such a model should allow economists to simultaneously represent both objective strategic considerations as well as agents' flawed interpretation thereof, and, critically, allow economists to construct such subjective interpretations from observable data.

This paper provides such a model, presenting a tractable framework to analyze a decision maker's hypothetical judgements. Using this model, I show that a DM's propensity to engage in hypothetical thinking is captured exactly by her ability to recognize implications (i.e., to identify that one hypothesis implies another) and that this later relation is identified by a DM's observable behavior. Thus, this characterization both provides a concrete definition of (flawed) hypothetical reasoning and, importantly, yields a methodology to identify these judgments from standard economic data.

If $\W$ is a state space representing all relevant resolutions of uncertainty, then a hypothesis is a collection of contingencies $H\subseteq \W$; if uncertainty resolves so that the true state-of-affairs in contained in the collection $H$, then the hypothesis is true, otherwise it is false. 
Therefore, the DM's flawed judgement is that she acts as if the hypothesis $H$ was instead some other hypothesis $\pi(H)\subseteq \W$; I call $\pi(H)$ her \emph{interpretation} of $H$ (and the map $\pi$, an interpretation). 

%A classic but oversimple example: a bidder falling victim to the \emph{winner's curse} \citep{thaler1988anomalies} in a common value auction. The bidder fails to recognize that if she wins, her signal was likely the highest among all bidders. Thus, her flawed interpretation of the hypothesis ``my bid wins the auction'' erroneously includes the possibility that her signal was central in the distribution of realized signals. Under this misinterpretation, she places too much probability on the contingency where she wins the auction \emph{and} her signal was accurate and she consequently overvalues the item.

%A classic but oversimple example: a restaurant receives a hygiene rating of `Pass' or `Fail' and then voluntarily discloses it by posting the grade in the window. Of course, all passing restaurants will post their grade, indicating to rational agents that a restaurant without a rating has failed the inspection. 
%Agents, however, do not seem to be rational, and in general will fail to extract all information from the absence of a signal \citep{jin2015no, enke2020you}---it seems that agents correctly interpret the revelation of `Pass', but fail to rule out the contingency that the restaurant passed following the lack of disclosure yielding $\pi(\text{`Pass'}) = \text{`Pass'}$ and $\pi(\text{`Fail'}) = \{\text{`Pass'}, \text{`Fail'}\}$.

A \emph{hypothetical expected utility} (HEU) maximizer values contingent claims (i.e., act from $\W$ to utils) according to both her probabilistic assessment of $\W$ as well as her interpretation of the contingencies on which the act depends. A HEU DM is therefore given by a measure on the state space, $\mu$, and a coherent interpretation, $\pi: 2^\W \to 2^\W$. Then, the DM values a state-contingent act $f: \W \to \R$ as
\begin{align}
\tag{\textsc{heu}}
V(f) =  \int^\mathscr{C} f \ \textup{d}(\mu\circ \pi),
\end{align}
where the integral is the Choquet integral, since in general $\mu\circ \pi$ will not be additive. Let $b_H$ denote the act which pays 1 util if $H$ is true and 0 otherwise. The HEU value of $b_H$ is therefore not the $\mu$-probability of $H$, but the $\mu$-probability of the DM's \emph{interpretation} of $H$,  $\mu(\pi(H))$. When $\pi$ is the identity, so the DM is a perfect hypothetical reasoner, HEU reduces to subjective expected utility. 

In the sequel, I show that HEU is characterized by restrictions on preference and that the parameters $\mu$, and more importantly, $\pi$, are identified. Properties of $\pi$ are reflected naturally in conditions on observable behavior. Thus, from preference, a modeler can understand how the DM interprets hypotheses and the limits of her cognitive abilities.

I establish these characterization and identification results for $\pi$ by considering \emph{subjective implication}: the DM's subjective assessment that one hypothesis implies another. Hypothetical thinking is intimately related to---indeed, in the present formulation, equivalent to---the ability to recognize implications. In other words, I show that a DM's subjective implications expose exactly what she envisions when presented with a hypothesis and thus her hypothetical reasoning. The HEU model, therefore, helps to make sense of \emph{why} hypothetical reasoning fails, by identifying which implications are misunderstood by the DM.

Importantly, I put forth a behavioral definition of subjective implication and show how it is possible to identify subjective implication from observed betting behavior. This `closes the model' so that the equivalent notions of subjective implication and the interpretation of hypotheses, $\pi$, are identified from empirically relevant data.  

%(and how it related to the implications perceived by a DM):

The rest of the paper is structured as follows: The next section details how HEU can make sense of several well known empirical failures of hypothetical thinking. It also discusses the characterization and identification results in more detail. Section \ref{sec:imp} introduces the notion of an interpretation map its relevant properties. It also introduces subjective implication and provides the correspondence between these two structures.  Then, Section \ref{section:model} provides the behavioral counterpart to the theory, \emph{Hypothetical Expected Utility}, using a DM's preference over contingent claims. An axiomatization of Hypothetical Expected Utility, in terms of observable restrictions on preference is provided in Section \ref{section:axioms}. Section \ref{sec:aa} discusses the relation between hypothetical reasoning and attitudes towards ambiguity. A brief survey of the relevant literature is found in Section \ref{sec:lit}. Proofs are located in the appendix. 

\section{Examples and Discussion of Results}

To capture the notion that the DM, while failing to precisely conceptualize the hypothesis, is otherwise rational, I consider \emph{coherent} interpretations which satisfy three properties. First, \emph{truth} requires $H \subseteq \pi(H)$ or that the DM's interpretation of a hypothesis will never rule out any contingencies that are, in reality, compatible with $H$, although she may err in the opposite direction;\footnote{Section \ref{sec:aa} considers the dual model, wherein the DM interprets each hypothesis $H$ as a \emph{smaller} set of contingencies: $\pi(H) \subseteq H$.} 
second, \emph{introspection} requires $\pi(\pi(H)) = \pi(H)$ or that the DM cannot distinguish between a hypothesis and her interpretation of it---if this were false then she should be be able to use this information to deduce that she is misperceiving $H$ in the first place;
finally, \emph{distribution} requires $\pi(H\cup G) = \pi(H)\cup \pi(G)$.
Distribution ensures that the DM is minimally consistent: that, in a weak sense, she believes the implications of her own beliefs. In particular, if the DM perceives the contingency $\w$ to be inconsistent with both $H$ and $G$ (so $\w\notin \pi(H)$ and $\w \notin \pi(G)$) then she can exclude $\w$ by knowing that either $H$ or $G$ holds without knowing which ($\w \notin \pi(H \cup G)$). Likewise, if the DM perceives the contingency $\w$ to be consistent with either $H$ or $G$, then she will not rule it out without knowing which of $H$ or $G$ holds. 

In each of the following examples---the monty hall paradox, the pivotal voting, the winner's curse, and disclosure games---HEU, with a coherent interpretation, can explain the common empirical behavior attributed to flawed hypothetical reasoning.
 
\begin{ex}{The Monty Hall Paradox}
\label{ex:mh}
At then end of the game show \emph{Let's Make a Deal}, the contestant is faced with three doors. Behind two of the doors sits a goat and behind the third, a prize. The contestant is asked to choose a door, after which the host of the show, Monty Hall, opens one of the unchosen doors and reveals a goat. Crucially, Monty himself knows the contents of the doors and \emph{always} opens a door with a goat behind it.  

The contestant is then afforded a final choice: stick with her chosen door or switch to the other (closed) door. Assuming all randomizations (i.e., the allocation of the prize and Monty's revelation) were uniform,\footnote{And, perhaps more controversially, assuming that the contestant prefers the prize to a lovely pet goat.} what should the contestant do? We can analyze the uncertainty faced by the contestant using a simple 4-state model.

By possibly relabelling the doors, we can assume without loss of generality that the contestant initially selects door 1. Then the 4 relevant states are:
$\W = \{\w_{12}, \w_{13},\w_{23},\w_{32}\}$, where $\w_{ij}$ represents the state where the prize is hidden behind door $i$ and Monty opens door $j$. Let $P_i$ denote the event that the prize is behind door $i$ and $O_j$ the event that Monty opens door $j$. These states are shown in Figure \ref{fig:MH}. Our assumption of uniform randomization indicates the probability over the four states (in the same order as listed above) is $\mu = [\frac16,\frac16,\frac13,\frac13]$. 

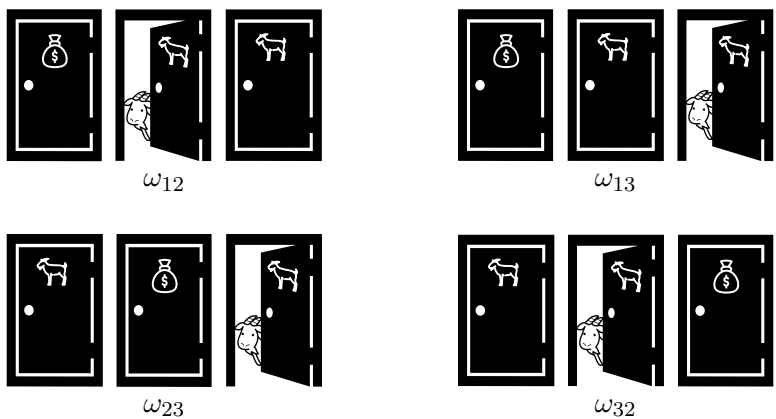
\begin{figure*}[]
\centering
\begin{tikzpicture}
\node (w12) at (0,3) {\includegraphics[width=.3\textwidth]{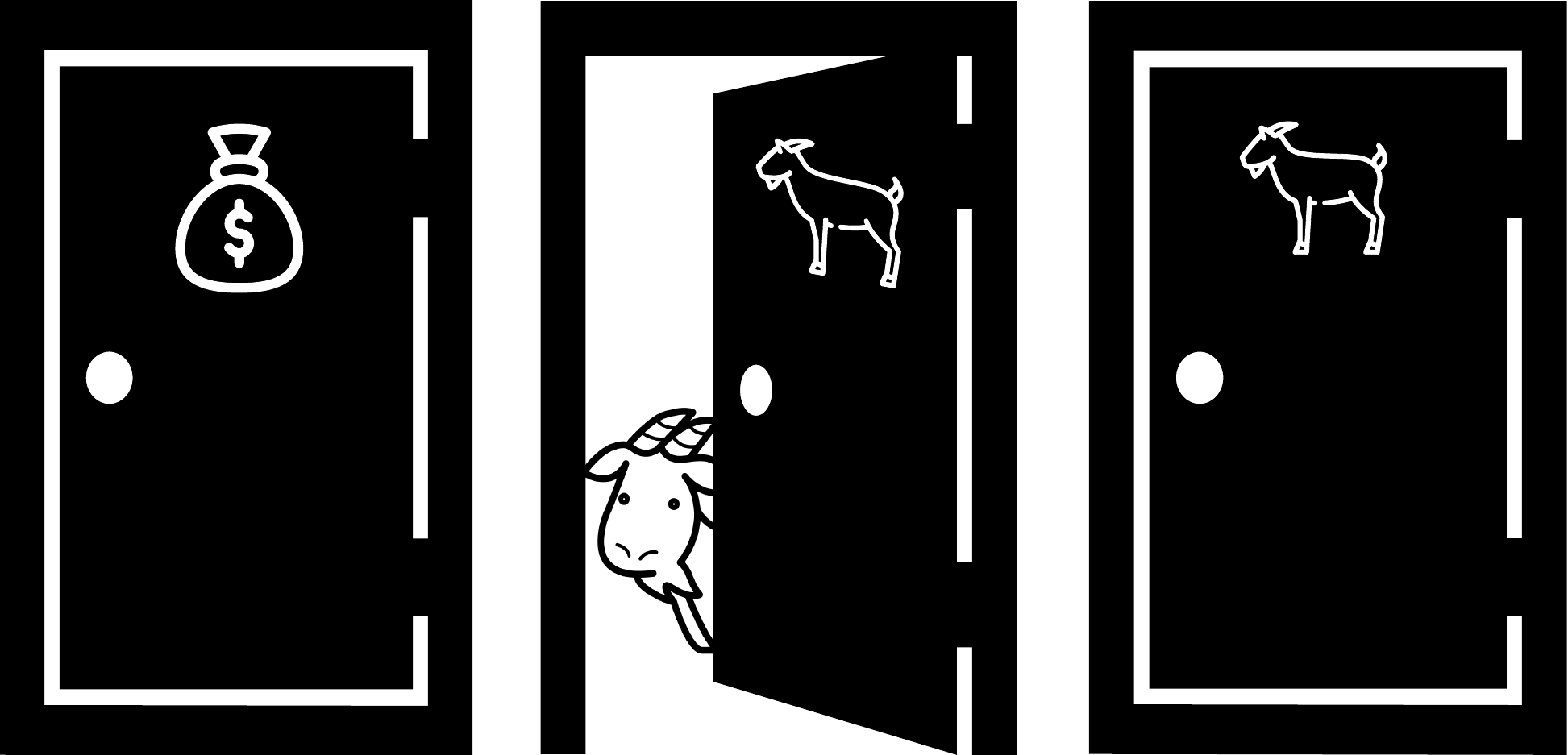}};
\node [below= -.1cm of w12] {$\w_{12}$};
\node (w13) at (6,3) {\includegraphics[width=.3\textwidth]{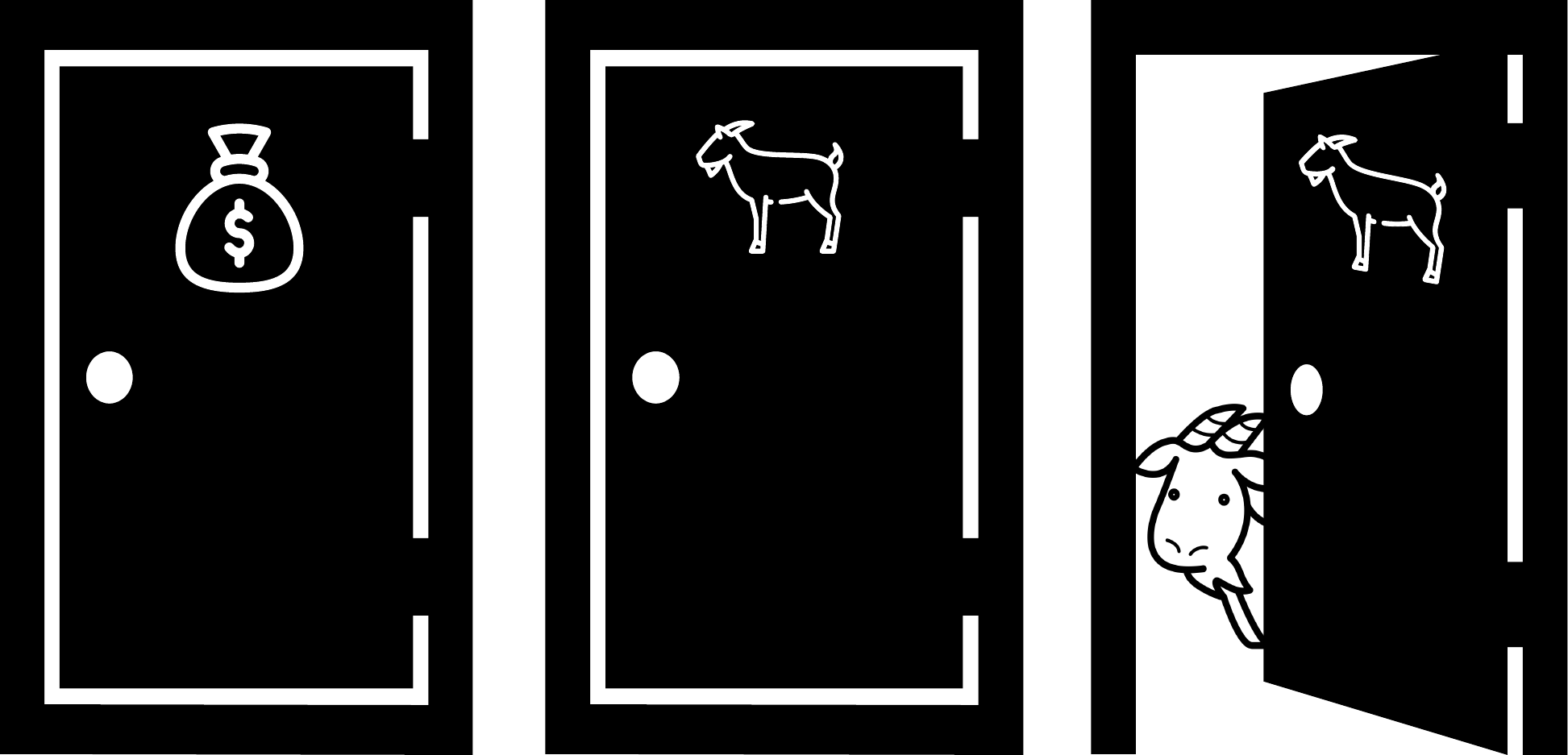}};
\node [below= -.1cm of w13] {$\w_{13}$};

\node (w23) at (0,0) {\includegraphics[width=.3\textwidth]{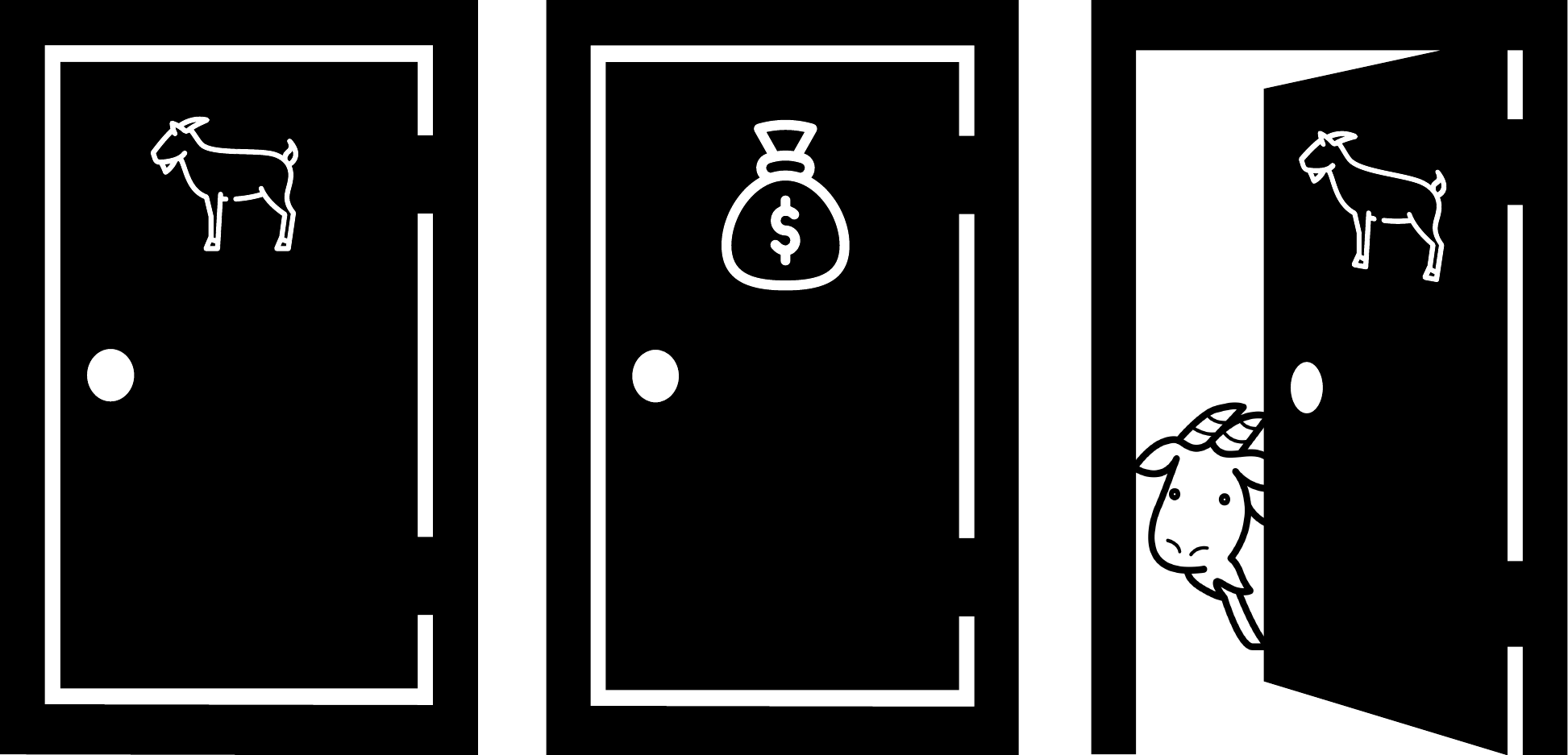}};
\node [below= -.1cm of w23] {$\w_{23}$};
\node (w32) at (6,0) {\includegraphics[width=.3\textwidth]{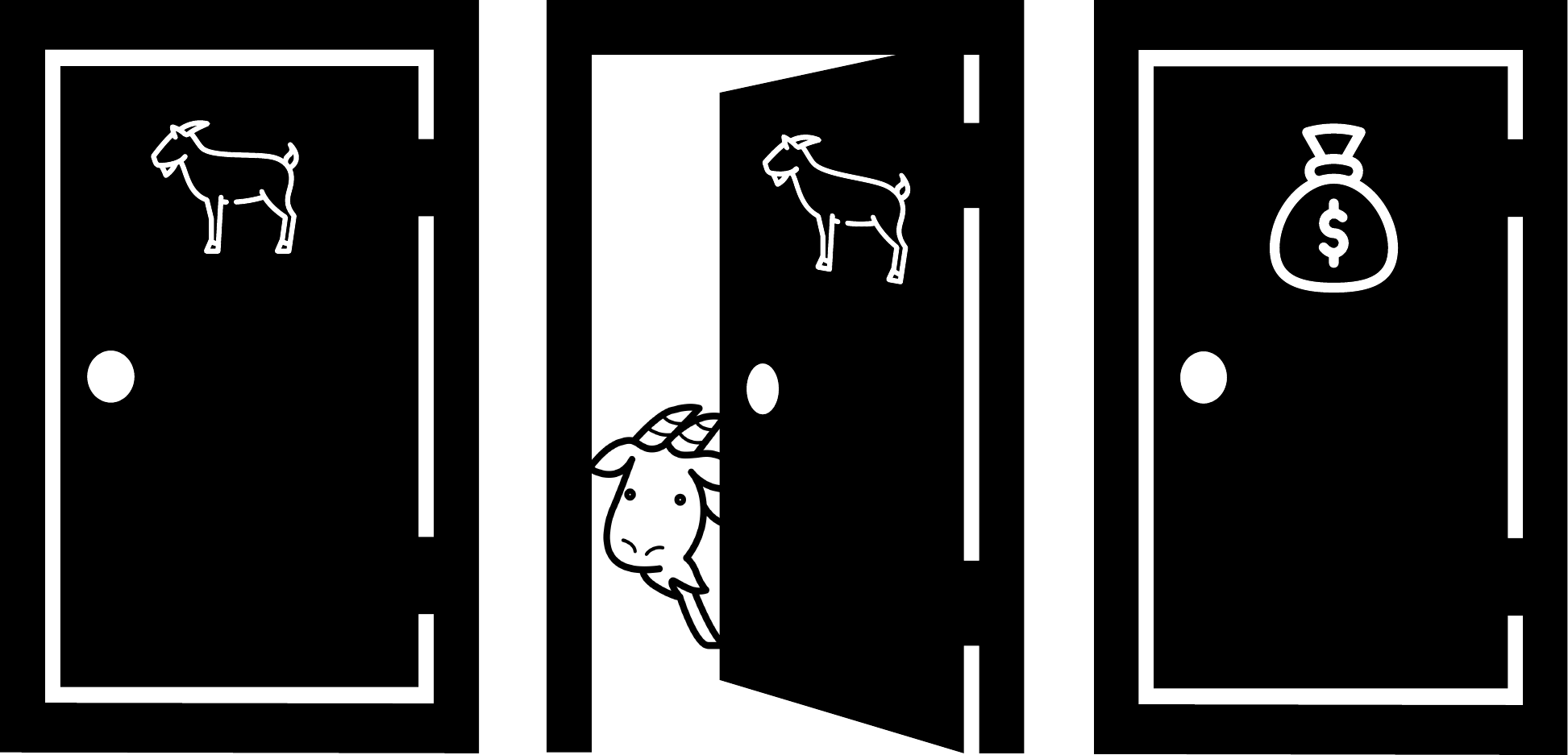}};
\node [below= -.1cm of w32] {$\w_{32}$};
\end{tikzpicture}
   \caption{States in the Monty Hall Problem. Of course, the symbols on the doors indicating their contents are for our benefit and are invisible to the contestant when making her decision.} 
\label{fig:MH}
\end{figure*}

Assuming the prize provides 1 util and losing 0, then the (objective) value of betting on door 1 after Monty opens door 2, is
$$V(b_{P_1} \mid O_2) = \mu(P_1 \mid O_2) = \tfrac{\mu(P_1 \cap O_2)}{\mu(O_2)}=  \tfrac{\mu(\{\w_{12}\})}{\mu(\{\w_{12}, \w_{32}\})}  = \frac{\tfrac{1}{6}}{\tfrac{1}{2}} = \frac13,$$ 
and the value of betting on door 3 is 
$$V(b_{P_3} \mid O_2) = \mu(P_3 \mid O_2) = \tfrac{\mu(P_3 \cap O_2)}{\mu(O_2)}  =  \tfrac{\mu(\{\w_{32}\})}{\mu(\{\w_{12}, \w_{32}\})}  = \frac{\tfrac{1}{3}}{\tfrac{1}{2}} = \frac23.$$ The same logic applies when Monty opens door 3 and so in every case, the contestant should (objectively) strictly prefer to switch doors.

Of course, contestants on the actual \emph{Let's Make a Deal} as well as countless subsequent thought experimenters have resisted this simple conclusion, even when presented with an analysis like the one above. Their flawed logic often follows: There is no initial bias between door 1 and door 3 as evidenced by $\mu(P_1) = \mu(P_3) = \frac13$, and Monty's revelation of a goat between door 2 does not directly concern the contents of doors 1 and 3, and so, even once door 2 has been eliminated the probability of winning is independent of the choice to switch.

This logic, although tempting, is specious. Its failure is precisely a failure of hypothetical reasoning---the contestant who does not strictly prefer to switch does so because she does not properly interpret the hypothesis $O_2$. Consider the interpretation $\pi: 2^\W \to 2^\W$ which is the identity on every hypothesis except $\pi(O_2) = \{\w_{12}, \w_{13},\w_{32}\}$ and $\pi(O_3) = \{\w_{12}, \w_{13},\w_{23}\}$. It is easy to check that $\pi$ is coherent.
Here, the contestant interprets ``Monty opens door 2'' as the event ``The prize is not behind door 2.'' While plausible, this is incorrect---it fails to incorporate the additional information contained in the event $O_2$, namely, that Monty could have chosen door 3 if prize is behind door 1, but this did not happen.

Using the same probabilistic beliefs but under this new interpretation:
\begin{align*}
V(b_{P_1} \mid O_2)  = \mu(\pi(P_1) \mid \pi(O_2)) &=  \tfrac{\mu(\{\w_{12},\w_{13}\})}{\mu(\{\w_{12},\w_{13},\w_{32}\})}  = \frac{\tfrac{2}{6}}{\tfrac{2}{3}} = \frac12, \text{ and} \\
V(b_{P_3} \mid O_2)  = \mu(\pi(P_3) \mid \pi(O_2)) &=  \tfrac{\mu(\{\w_{32}\})}{\mu(\{\w_{12},\w_{13},\w_{32}\})}  = \frac{\tfrac{1}{3}}{\tfrac{2}{3}} = \frac12,
\end{align*}
the HEU contestant indeed sees no benefit to switching doors.
\end{ex}

%To capture the notion that the DM, while failing to precisely conceptualize the hypothesis, is otherwise rational, I consider \emph{coherent} interpretations. An interpretation $\pi$ is coherent if it satisfies, in addition to $\pi(H) \subseteq H$ it satisfies two properties.
%The first, \emph{introspection} requires $\pi(\pi(H)) = \pi(H)$.
%This requirement is tantamount to assuming a DM cannot distinguish between a hypothesis $H$ and her interpretation of it, $\pi(H)$---indeed, if this were false then she should be be able to use this information to deduce that she is misperceiving $H$ in the first place. As such this restrictions has a flavor of KU introspection from \cite{dekel1998standard}, which requires that a DM never know that she is unaware of an event. 
%
%The second additional restriction, \emph{distribution} requires $\pi(H\cup G) = \pi(H)\cup \pi(G)$.
%Distribution ensures that the DM is minimally consistent, that, in a weak sense, she believes the implications of her own beliefs. In particular, if the DM perceives the contingency $\w$ to be inconsistent with both $H$ and $G$ (so $\w\notin \pi(H)$ and $\w \notin \pi(G)$) then she can exclude $\w$ by knowing that either $H$ or $G$ holds without knowing which. Likewise, if the DM perceives the contingency $\w$ to be consistent with either $H$ or $G$ (so $\w\in \pi(H)$ or $\w \in \pi(G)$) then she will not rule it out without knowing which of $H$ or $G$ holds. 

\begin{ex}{Pivotal Voting}
\label{ex:ev}
\def\B{\textcolor{gray!70!blue}{\textbf{B}}}
\def\b{\textcolor{gray!70!blue}{\textbf{b}}}
\def\R{\textcolor{gray!50!red}{\textbf{R}}}
\def\r{\textcolor{gray!50!red}{\textbf{r}}}
\def\p{\textsc{p}}
The fact that the voter is decisive is informative about how the rest of the electorate voted, and therefore, may be indirectly informative about the value of the alternatives being voted for. A rational voter should condition on the contingency that she is pivotal, as in any other state of affairs her action is inconsequential. \cite{esponda2014hypothetical} convincingly demonstrate that even in very simple (non-strategic!) experimental setting subjects do not condition correctly and consequently  play dominated strategies. Consider the following simplification of their experiment: 

\begin{quotation}
The state of the world is either \R\ or \B\ with equal probability. The  subject  must  cast  a  vote  for either \R\ or \B\ without observing the state but after observing a signal---the possible signal realization are \r\ and \b, with accuracy $\frac23$.\footnote{Specifically, \r\ realizes with probability $\frac23$ in state \R\ and $\frac13$ in state \B\ and \b\ with the remaining probably: $\frac13$ in state \R\ and $\frac23$ in state \B.} In addition, two computers observe the state and are programmed to follow specific rules for casting a vote in favor of \R\ or \B\ contingent on the (realized) state. If a simple majority votes for the correct state (i.e., the votes match the state), the subject's payoff is \$2; otherwise, the payoff is \$0. 
Before casting her vote, the subject receives information about the rule being followed by the computers, but does receive information about the actual votes of the computers. Both computers follow the rule: (i) If the state is \R: vote \R; (ii) If the state is \B: vote \B\ with probability $\frac23$ and \R\ with $\frac13$. 
\end{quotation}

Notice that voting \B\ is a (weakly) dominant strategy since in order for a subject's vote affect the outcome, the computers must disagree, and hence the state must be \B.  Nevertheless, 80\% of subjects do not play strategically even after 40 rounds of play.\footnote{This was defined as voting \textsc{red} more than 15\% of the time.} 

The HEU model can represent the subjects' inability to think hypothetically, specifically, that the subjects do not correctly understand the logical relationship between the state and the possibility of being pivotal. 

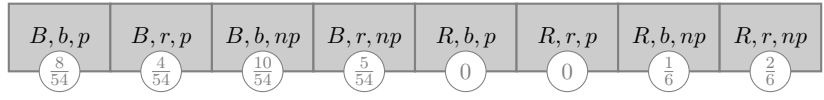
\begin{figure}
\centering

\begin{tikzpicture}[scale=.9, every node/.style={scale=.8}, level/.style={},]

\coordinate (e1) at (0,1);
\coordinate (e2) at (4.5,1);

\coordinate (f1) at (1.5,-2.5);
\coordinate (f2) at (4.5,-2.5);

\draw[gray, fill=gray!40, thick] (0,0) rectangle (1.5,1) node[pos=.5, color=black] {$B,b,p$};
\draw[gray, fill=gray!40, thick] (1.5,0) rectangle (3,1) node[pos=.5, color=black] {$B, r, p$};
\draw[gray, fill=gray!40, thick] (3,0) rectangle (4.5,1) node[pos=.5, color=black] {$B,b,np$};
\draw[gray, fill=gray!40, thick] (4.5,0) rectangle (6,1) node[pos=.5, color=black] {$B,r,np$};
\draw[gray, fill=gray!40, thick] (6,0) rectangle (7.5,1) node[pos=.5, color=black] {$R,b,p$};
\draw[gray, fill=gray!40, thick] (7.5,0) rectangle (9,1) node[pos=.5, color=black] {$R,r,p$};
\draw[gray, fill=gray!40, thick] (9,0) rectangle (10.5,1) node[pos=.5, color=black] {$R,b,np$};
\draw[gray, fill=gray!40, thick] (10.5,0) rectangle (12,1) node[pos=.5, color=black] {$R,r,np$};

\draw[gray,fill=white] ($(0,0)!0.5!(1.5,0)$) circle (.3) node{$\tfrac{8}{\small{54}}$};
\draw[gray,fill=white] ($(1.5,0)!0.5!(3,0)$) circle (.3) node{$\tfrac{4}{\small{54}}$};
\draw[gray,fill=white] ($(3,0)!0.5!(4.5,0)$) circle (.3) node{$\tfrac{10}{\small{54}}$};
\draw[gray,fill=white] ($(4.5,0)!0.5!(6,0)$) circle (.3) node{$\tfrac{5}{\small{54}}$};
\draw[gray,fill=white] ($(6,0)!0.5!(7.5,0)$) circle (.3) node{$0$};
\draw[gray,fill=white] ($(7.5,0)!0.5!(9,0)$) circle (.3) node{$0$};
\draw[gray,fill=white] ($(9,0)!0.5!(10.5,0)$) circle (.3) node{$\tfrac16$};
\draw[gray,fill=white] ($(10.5,0)!0.5!(12,0)$) circle (.3) node{$\tfrac26$};

\end{tikzpicture}
\caption{The probabilities for the 8 states in Example \ref{ex:ev}.}
\label{fig:ex2}
\end{figure}

Let $\W = \{B,R\} \times \{b,r\} \times \{p,np\}$. Each state therefore represents the realization of three uncertainties: $\{B,R\}$ encodes the state being \B\ or \R, $\{B,R\}$ encodes the realized signal being \b\ or \r, and $\{n,np\}$ encodes whether, given the computers votes, the subject is pivotal or not. The probabilities of each state, as given by the rules governing the realization of uncertainty as laid about above, are given by Figure \ref{fig:ex2}. 

The purely rational subject, whose interpretation would be given by the identity, would indeed see no value in voting \R. To see this notice, the relevant conditioning event is that the subject observed the \r\ signal and can affect the outcome: $\{B,R\} \times \{r\} \times \{p\}$ which we refer to in an abuse of notation as $\{r,p\}$. Conditional on this event, \B\ is indeed more likely than \R---$\mu(B \mid \{r,p\}) = 1 > 0 = \mu(R \mid \{r,p\})$---so the subject is better of ignoring her private signal. 

Of course, the subjects were not perfectly rational, and indeed, fail to condition on being pivotal. This failure can be seen through the lens of a flawed interpretation: let $\pi$ be the interpretation map that ignores pivotality. Specifically: define $\pi$ over $\W$ via
$$ \pi(S,s,p) = \pi(S,s,np) = \{(S,s,p),(S,s,np)\},$$  
for $(S,s) \in \{B,R\} \times \{b,r\}$, and extend $\pi$ to $2^\W$ via $\pi(E) = \bigcup_{\w\in E} \pi(\w)$.
so that $\{(B,b,p)\} \mapsto \{(B,b,p), (B,b,np)\}$ and $\{(B,r,p)\} \mapsto \{(B,r,p), (B,r,np)\}$ and so forth. It is easy to check that $\pi$ is coherent. Such a subject does not distinguish between events which differ only insofar as pivotality. Hence, when considering the optimal action after observing \r, the subject erroneously bases her decision on possibility that the state is $\R$. Specifically, rather than conditioning on $\{r,p\}$, the subject, whose view of the world is unconcerned with pivotality, conditions on $\pi(\{r,p\}) = \{B,R\} \times \{r\} \times \{p,np\}$. In line with the observed subjects' behavior, we have $\mu(R \mid \pi(\{r,p\})) = \frac23 > \frac13 = \mu(B \mid \{r,p\})$.
\end{ex}

\begin{ex}{The Winner's Curse}
\label{ex:wc}
The winner's curse, the phenomenon whereby bidders in a common value auction fail to condition on having received the most optimistic private signal and hence overbid, can be captured in a very similar fashion as to Example \ref{ex:ev}.  To see this, consider a stylized version of the problem wherein the object for auction has a common value, taking values on some set $V$. Each bidder $i = 1 \ldots n$ receives a private signal $s \in S$ correlated with the value of object. 

After observing $s$ how should a rational bidder bid? She should calculate her value conditional on $s$ \emph{and} the fact that she wins the auction, since her bid is irrelevant any other time. We can model this, like the previous example, with with a state-space that is the product space of three sources of uncertainty $\W = V\times S \times V$: for $(v,s,u) \in \W$, $v$ is the value of the object, $s$ the private signal of the bidder and $u$ the upper-bound of all other bidders' bids. The probability $\mu$ depends on the distribution of value, the signal generation process, and  the strategic model in the bidder's head (which in an equilibrium would be the other bidder's true behavior). Hence the optimal bid, $b$, must be based on maintaining a value $\mathbb{E}(v \mid V \times \{s\} \times \{u | u > b\})$. 

To capture the winner's curse, we must represent the bidder's flawed hypothetical reasoning. In particular, her propensity to ignore the fact that the winning the auction is informative about other bidders' private information and hence the value of the object itself. As above, ignoring this fact is tantamount to ignoring a dimension of the state-space: define $\pi$ over $\W$ via
$$ \pi(v,s,u) = \bigcup_{u' \in V} (v,s,u')$$  
for $(v,s) \in V \times S$, and extend $\pi$ to $2^\W$ via $\pi(E) = \bigcup_{\w\in E} \pi(\w)$. Now, the bidder, after seeing signal $s$ maintains a expected value for the object $\mathbb{E}(v \mid \pi(V \times \{s\} \times \{u | u > b\})) = \mathbb{E}(v \mid V \times \{s\} \times V)$ which is independent of the other bidders' strategies. 
\end{ex}

\begin{ex}{Disclosure Games}
\label{ex:dg}
Strategic concerns often ensure that rational agents will voluntary disclose their information, because in equilibrium, no news will interpreted as bad news. As a classic example, a seller can credibly disclose the value of the object for sale, taking integer values from 1 (lowest) to $n$ (highest) with equal probability. Surely, if object has the highest value, $n$, the seller would display the value, and so a rational prospective buyer who sees a seller without any posted valuation can assume the value must be $n-1$ or less. But then, sellers with objects of value $n-1$ would want their value to be known, rather than letting the buyer be uncertain of their position but knowing it is less than $n-1$.  Continuing inductively, it follows that all types of seller's (except possibly the lowest type) will make public the value of their good.  

Of course, this is not the case in real life manifestations of the above scenario. In an experimental setting, \cite{jin2015no} find

\begin{quotation}
``[Sellers] disclose favorable information, but withhold less favorable information. The degree to which [sellers] withhold information is strongly related to their stated beliefs about [buyer] actions, and their stated beliefs are accurate on average. ... [Buyer] actions and beliefs suggest they are insufficiently skeptical about non-disclosed information in the absence of repeated feedback. "
\end{quotation}

Again, this can be straightforwardly captured through the HEU model. Let $\W = \{v_1 \ldots v_n\} \times \{r_0, r_1 \ldots r_n\}$, where the $(v_i,r_j)$, $j \neq 0$, corresponds to state in which the value of the object is $i$ and the seller reveals the object has value $j$ and $(v_i,r_0)$ to the state where the value is $i$ and the seller does not reveal anything. Let $V_i = \{v_i\} \times  \{\emptyset, r_1 \ldots r_n\}$ and $R_i =  \{v_1 \ldots v_n\} \times \{r_i\}$.

Let the sellers strategy be encoded by $\beta_1 \ldots \beta_n$, where $\beta_i$ is the probability that a seller of quality $i$ reveals his type. Then the probability of the states is given by $\mu$ where $\mu((v_i,r_i)) = \frac1n\beta_i$, $\mu((v_i,r_0)) = \frac1n(1-\beta_i)$ and, since sellers can only reveal truthful information, $\mu((v_i,r_j)) = 0$ for $i \neq j \neq 0$.

Notice that if $\beta_n = ... = \beta_{n-k} = 1$---the highest $k$ types always reveals---then for the rational buyer $\mu(V_n \cup ... \cup V_{n-k} \mid R_0) = 0$. Upon seeing no revelation, the rational buyer understands the seller is not of the $k$ highest types, and in particular, that the expected quality is strictly less than $n-k$. This is critical to maintain the unraveling argument outlined above. 

Consider instead a buyer with constrained hypothetical reasoning, given by 
\begin{equation}
\pi: A \mapsto \begin{cases}
A \text{ if } A \cap R_0 = \emptyset \\
\W \text{ otherwise}.
\end{cases}
\end{equation}

As always $\pi$ is coherent. Such a buyer correctly understands that disclosure is truthful, but treats no-disclosure as if no information had been revealed (i.e., she fails to extract information from the \emph{lack} of a signal). Indeed, upon observing $R_0$, the buyer does not update her beliefs about quality at all: $\mu(V_n | R_0) = \mu(V_n) = \frac1n$. Hence for lower-than-average quality types, independent of other types equilibrium strategies, it is better to not reveal anything so as to maintain the average (prior) belief of the buyer.
\end{ex}

\subsection{Subjective Implication}

Understanding a hypothesis, $H$, is understanding what is implied by, and what implies, $H$. A hypothesis $H$ implies a hypothesis $G$ if whenever $H$ is true, $G$ is as well: more formally, when $H \subseteq G$. A DM with interpretation $\pi$ \emph{perceives} an implication $H \implies G$, not according to their true relation but according to the interpretation: the DM perceives $H \implies G$ iff $\pi(H) \subseteq \pi(G)$.
A DM who fails to extract all information from $H$ will perceive too many implications:\footnote{It may seems initially surprising that a the DM perceives too many, rather than too few, implications. However, an over perception of implication is precisely the effect of conflating distinct events, itself the product of flawed hypothetical reasoning. For example, if $H$ is objectively a strict subset of $G$ but the DM fails to extract all information from the hypothesis $H$, instead interpreting it as identical to $G$, then she will perceive the erroneous implication $G \implies H$.} in Example \ref{ex:mh}, the contestant interprets $O_2$ (door 2 is opened) as $\n(P_2)$ (the prize is not behind door 2; i.e., $\W \setminus P_2$). Thus, she correctly perceives $O_2 \implies \n(P_2)$, since $\pi(O_2) \subseteq \pi(\n(P_2))$, but also incorrectly perceives $\n(P_2) \implies O_2$, since $\pi(\n(P_2)) \subseteq \pi(O_2)$. 

In this paper, I present a model of \emph{subjective implication},  assuming a modeler can observe the DM's perceived implications, given by a binary relation $\implies$ over the set of hypotheses. I provide restrictions on $\implies$ to ensure the existence of a coherent $\pi$, such that $H \implies G$ if and only if $\pi(H) \subseteq \pi(G)$. Moreover, this interpretation is unique and can be found constructively.

Beyond uncontroversial transitivity and monotonicity type restrictions, coherency is captured by two conditions on $\implies$. First, the DM must be able to engage in \emph{deduction}: 
If the DM perceives that both $H$ and $H'$ would imply $G$ she does not need to know which of $H$ and $H'$ hold in order to draw a conclusion. 
Second, the DM can \emph{decompose} complex implications into simpler ones. In particular, if $F$ implies that either $H$ or $H'$ is true, but does not inform the DM as to which, then $F$ can be decomposed into the disjunction two stronger hypotheses, $G$ and $G'$, which themselves imply $H$ and $H'$ respectively.\footnote{Formally, deduction is `$H \implies G$ and $H' \implies G$ then $H \cup H' \implies G$,' and decomposition is `$F \implies H \cup H'$ implies there exists some $G, G'$, such that $G \implies H$, $G' \implies H'$ and $F = G \cup G'$.'}
This second condition requires some examination. 

Consider a hypothesis $F$ which implies either $H$ = `the student will do very well on the final exam' or $H'$ = `the student will do very poorly on the final exam' but does not determine which is true. Then decomposition requires that $F$ can be decomposed into two sub-hypotheses which imply $H$ and $H'$ independently. Perhaps $F$ = `the student left the 2-hour final exam after 20 minutes.' Then $F$ can be itself decomposed into `the student is very bright and understands the material completely' which implies $H$ and `the student is apathetic and does not care about her grade in the class' which implies $H'$.

%and provide the conditions under which it is possible, from a DM's perceived implications, to reconstruct her interpretation of hypothetical events, and hence her hypothetical reasoning. If a modeler (he) observed that the DM (she) believes that $H$ implies $G$ then he could conclude that her interpretation of $H$ is a subset of her interpretation of $G$: $\pi(H) \subseteq \pi(G)$. I show how observing subjective implication can be used to identify $\pi$, what a DM envisions when presented with a hypothesis, and explore how conditions on perceived implication manifest as conditions on hypothetical reasoning. 

It is not immediately clear that the subjective implication relation is empirically meaningful. To allay such worries, I provide a methodology for identifying, from observable behavior, which implications a DM perceives, and therefore, her power to reason hypothetically. As is standard, a DM's choice over contingent payoffs identifies her beliefs about the events over which the acts are contingent. For example, if $b_H$ is a bet on the hypothesis $H$---a contingent claim that pays if and only if $H$ is true---then the DM's preference to bet on $b_H$ rather than $b_G$ would expose her belief that the $H$ is the more likely of the two hypotheses. A similar idea identifies not only the ordinal likelihood of hypotheses but also their implications.

Because the contestant in the example mistakenly conflates $O_2$ with $\n(P_2)$, she does not conceive of the possibility that $\n(P_2)$ could be true while $O_2$ is false; indeed, this is what it means to perceive $\n(P_2) \implies O_2$. This presents as her indifference between $b_{O_2}$ and $b_{O_2 \cup \n(P_2)}$. In other words, the additional possibility of winning contingent on $\n(P_2)$ is of no value, as the contestant believes that whenever $\n(P_2)$ is true, so too is $O_2$, and therefore, she would win the bet regardless. 
Conversely, starting with preferences over acts, I take a DM's indifference between $b_{G}$ and $b_{G\cup H}$ as the definition of her perceiving that $H$ implies $G$. Thus, from observed preference, it is possible to construct a notion of subjective implication and subsequently of hypothetical interpretations.\footnote{That this identification is consistent rests on sought after hypothetical interpretation $\pi$ (or equivalently, on the subjective implication relation) being coherent, as embodied by axiomatic restrictions on preference.} 
Exploiting the preference based definition of subjective implication, I provide an axiomatic characterization of HEU in the case where $\pi$ is coherent. In such a case, the parameters of the model are identified. Thus, from preferences over contingent claims, the modeler can identify the DM's interpretation of hypothetical events, $\pi$, and her implied probabilistic beliefs over the true state space $\mu$.

In the example, the source of the contestant's indifference between switching or not---her misidentification of $O_2$ as $\n(P_2)$---also produces an affinity for ambiguity. Take the bets $b_{O_2}$ and $b_{O_3}$ which pay a prize of 1 on the given event and 0 otherwise. If the contestant is an HEU maximizer, then she values these acts both at $\mu(\pi(O_2)) = \mu(\pi(O_3)) = \frac23$. An equal mixture of these two bets, $\frac12b_{O_2} + \frac12b_{O_3} = \frac12b_{O_2 \cup O_3} = \frac12 b_\W$, however, is valued at $\frac12$. Thus the contestant displays an aversion to hedging. Below, I show this behavior is general: the HEU maximizing DM displays ambiguity seeking behavior. In a round about way, this helps make sense of winner's curse type overbidding and, more generally, over confidence in the face of hard to comprehend decision problems: by failing to exclude certain possibilities, DM's behave as if they have an affinity for actions with ambiguous or uncertain payoffs.

\section{Interpretation and Implication}
\label{sec:imp}

Let $\W$ be a set, thought of as the objective space of uncertainty relevant to a decision problem. Elements of $\W$ are called states or contingencies. A collection of contingencies (i.e., a subset of $\W$) is referred to as a hypothesis. When $H \subseteq G$ we state that $H$ is a stronger hypothesis than $G$ and $G$ is weaker than $H$. For two hypotheses, $H$ and $G$, call $H \cap G$ the \emph{joint} hypothesis. Note that these descriptions relate to the objective character of the hypotheses and not necessarily DM's perception of them. 

The decision maker might fail to properly perceive the hypothesis, $H$. That is, when thinking hypothetically, the DM misinterprets the hypothesis $H$ as some other event $H'$. We let  $\P: 2^\W \to 2^\W$, aptly called \emph{an interpretation}, represent the DM's interpretation of events.
 
Although the DM may fail to accurately identify events implied by a given hypothesis, we assume she is an otherwise competent reasoner as captured by the following two restrictions on $\P$:
\begin{itemize}[leftmargin=20ex]
\item[\mylabel{t}{T}{Truth}] $H \subseteq \P(H)$, 
\item[\mylabel{i}{I}{Introspection}] $\P(\P(H)) = \P(H)$,
\item[\mylabel{m}{M}{Monotonicity}] $H \subseteq G$ implies $\P(H) \subseteq \P(G)$, and
\item[\mylabel{c}{C}{ Consistency}] $\P(H\cup G) \subseteq \P(H) \cup \P(G)$.
\end{itemize}

The restriction \r{t} states that, while the DM may fail to extract the full information from the hypothesis $H$, and thus continue to entertain contingencies that should in fact be ruled out $H$, she is never deluded: she does not believe that $H$ eliminates those contingencies it is actually compatible with. The restriction \r{i} states that if the decision maker mistakenly perceives the hypothesis $H$ as $G = \P(H) \supset H$, then she perceives $G$ correctly. In other words, this ensures the DM cannot distinguish between a hypothesis and her interpretation of it. As such this \r{i} has a flavor of KU introspection from \cite{dekel1998standard}, which requires that a DM never know that she is unaware of an event.  \r{m} ensures the DM's interpretation does not \emph{reverse} strict implications. 

Finally, the restriction \r{c} states that the DM understands the implications of her own interpretation; i.e., that she is minimally consistent when contemplating different hypotheses. In particular, if she perceives a contingency is ruled out by the hypothesis $H$, and also by the hypothesis $G$, then she is able to rule out this contingency by knowing that either $H$ or $G$ is true without knowing which.

It is straightforward to show that interpretation $\pi$ satisfies \r{m} and \r{c} if and only if it satisfies:

\begin{itemize}[leftmargin=20ex]
\item[\mylabel{d}{D}{ Distribution}] $\P(H\cup G) = \P(H) \cup \P(G)$.
\end{itemize} 

Call an interpretation \emph{weakly coherent} if it satisfies \r{t}, \r{i}, and \r{m}, and call it \emph{coherent} if it satisfies, in addition, \r{c} (or equivalently if it satisfies \r{t}, \r{i}, and \r{d}).

%\begin{remark}
%If $\pi$ satisfies \r{c} then it is monotone:
%\begin{itemize}[leftmargin=20ex]
%\item[\mylabel{m}{M}{Monotonicity}] $H \subseteq G$ implies $\P(H) \subseteq \P(G)$.
%\end{itemize}
%Indeed, notice that if $H \subseteq G$  then $\P(H) \subseteq \P(H) \cup \P(G) = \P(H\cup G) = \P(G)$.
%\end{remark}

\subsection{Implication} 
Failures in hypothetical thinking can be thought of as failures in correctly assessing implications. A DM's misperception of a hypothesis $H$ is her failure to properly identify which contingencies imply $H$ is true, and which contingencies follow from the truth of $H$. 
We therefore begin the investigation in to a hypothetically challenged decision maker by considering the relation between her subjective notion of implication and the HEU representation. 

Assume for the moment that a modeler was able to identify a DM's `implication' relation. That is, a binary relation `${\implies}$' over hypotheses, with the interpretation that $H \implies G$ if and only if the DM perceives that the hypothesis $H$ implies the hypothesis $G$. Let $\d[H, \implies] =\{G \mid G \implies H\}$ collect the hypotheses that $\implies$ imply $H$. When the implication relation is clear from context, we will omit it, writing $\d[H]$, instead. At present, we take this relation as given, but rest easy, as subsequent sections will show that this data is encoded through traditional decision theoretic primitives. 

An implication relation, $\implies$, is \emph{derived from} an interpretation, $\pi$, if
\begin{equation}
\label{eq:derived}
\tag{\textsc{drv}}
H \implies G \text{ if and only if } \pi(H) \subseteq \pi(G)
\end{equation}

\begin{remark}
A clear necessary condition for $\implies$ to be derived from an interpretation is that it is reflexive and transitive. This is not, however, sufficient. The minimal additional sufficient conditions, while straightforward (i.e., combinatorial restrictions on the lengths of chains and anti-chains) does not lend itself to a sensible economic interpretation. 
\end{remark}

We are interested in a decision maker who may misinterpret some hypotheses but is otherwise rational in the sense that she understands the structure of her own knowledge. In other words, taken her misperception as given, her notion of implication obeys the standard tenets of logic: namely, that implication is transitive and that a joint hypothesis is implied whenever both of its components are. 

\begin{ax}{I}{trv}{transitivity}
If $H \implies G$ and $G \implies F$ then $H \implies F$.
\end{ax}

%\begin{ax}{I}{ded}{deduction}
%If $H \implies G$ and $H \implies F$ then $H \implies G \cap F$.
%\end{ax}

\begin{ax}{I}{ded}{deduction}
If $H_i \implies G$ for all $I \in \mathcal I$, then $\bigcup_{i \in \mathcal I} H_i \implies G$.
\end{ax}

In addition to these purely logical restrictions, we stipulate another restriction that relates the actual relation between hypotheses to the DM's interpretation of them. In particular, we state that a DM correctly understands the \emph{true} implication relation: in other words she correctly assesses when one hypotheses is weaker than another. 

\begin{ax}{I}{mon}{monotonicity}
If $H\subseteq G$ then $H \implies G$.
\end{ax}

From a particular vantage, this may seem like a strong assumption in that it leaves no room for the DM to fail to perceive implications. However, I argue that it is the \emph{over} perception of implication that leads to many common errors in hypothetical reasoning: indeed, to be able to distinguish between two events, is to perceive that one implies something the other does not (i.e., that $H \implies H$ while $G \centernot\implies H$). Thus, additional implications stem from the DM conflating events via misperception. 

\begin{proposition}
\label{prop:drvw}
If $\implies$ satisfies \textsc{\textbf{I1-3}}, if and only if it is derived from a weakly coherent interpretation. Moreover, this weakly coherent interpretation is unique.
\end{proposition}

A coherent interpretation imparts more structure. The (sub-)modularity of $\pi$ implies a implication of a disjunctive hypotheses can be decomposed into the disjunction of stronger hypotheses. Specifically:

\begin{ax}{I}{dcmp}{decomposition}
If $F \implies H \cup H'$ then there exists some $G, G'$, such that $G \implies H$, $G' \implies H'$ and $F = G \cup G'$.
\end{ax}

Consider a hypothesis $F$ that implies that either $H$ or $H'$ is true, but does not inform the DM as to which. In such a scenario, \rax{dcmp} states that $F$ can be decomposed into the disjunction two stronger hypotheses, $G$ and $G'$, which themselves imply $H$ and $H'$ respectively. As in the introduction, if a student leaving an exam very early is an indication that he is going to receive either an exceptionally  good or an exceptionally poor grade, then his leaving early can be decomposed into two stronger hypotheses: that he is very gifted or very apathetic.

\begin{proposition}
\label{prop:drv}
$\implies$ satisfies \textsc{\textbf{I1-4}}, if and only if it is derived from a coherent interpretation. Moreover, this coherent interpretation is unique. 
\end{proposition}

The uniqueness claims in the two prior propositions rely critically on the (weak) coherency of the interpretation: there may be other non-coherent interpretations from which $\implies$ is also derived. Nonetheless, under the assumption of coherency, Propositions \ref{prop:drvw} and \ref{prop:drv} establish a very tight connection between a DM's subjective implication and her interpretation of hypothetical events: given the observation of subjective implications it is possible to uniquely identify which contingencies the DM perceives when she contemplates a hypothesis. 

\section{An Observable Model}
\label{section:model}

\subsection{Notation}

In the sequel, I assume $\W$ is finite; the motivation for this restriction is explained in the Section \ref{sec:ifp}. The primitive of the model is a decision maker's preference over \emph{acts}. A act, $f: \W \to \R_+$ is a state-contingent payoff, in utils.\footnote{The assumption that payoffs are in utils is harmless given the body of decision theoretic results allowing for the conversion of lotteries or other alternatives into utils in a cardinally unique manner.} 
Let $\F$ collect all such acts and endow $\F$ with the with the topology of pointwise convergence.  The primitive of the model is a preference relation, $\s$, over $\F$.

For some $x \in \R_+$, we identify $x$ with the constant act which pays $x$ in every state. 
For two acts $f$ and $g$ let $f_{H}g$ denote the act which agrees with $f$ on $H$ and $G$ elsewhere. A \emph{bet} on $H$, denoted by $b_H$, is the act $1_H0$ and is which pays 1 on $H$ and 0 elsewhere. Let $f \l g$ denote the pointwise minimum between $f$ and $g$ and $\alpha f + (1-\alpha) g$ the pointwise $\alpha$ mixture between $f$ and $g$.

A \emph{capacity} on $\W$ is a function $\nu: 2^\W \to [0,1]$ such that $\nu(\emptyset) = 0$, $\nu(\W) = 1$ and $H \subseteq H'$ implies $\nu(H) \leq \nu(H')$. Notice that a probability measure is an additive capacity. Capacities carry a well defined notion of integration, that generalizes the usual notion of integration for measures. Given $f: \W \to \R_+$ and a capacity $\nu$, define the \emph{Choquet} integral to be
$$ \int^\mathscr{C} f \ \textup{d}\nu = \sum_{i=1}^{n} (x_i - x_{i-1})\nu(\{\w \mid f(\w) \geq x_i\}), $$
where $x_1 \ldots x_n$ are the values taken by $f$ in increasing order (i.e., $x_i \leq x_{i+1}$) and where $x_0$ is defined to be $0$.

\subsection{Representation}

%
%In addition, we assume that a DMs ability to identify events is better under a more specific hypothesis. For all $H, H'$ with $H' \subseteq H$ and $E \subset H'$
%
%\begin{itemize}
%\item[\mylabel{idp}{\textsc{idp}}{f}] $\P_{H}(E) = \P_{H'}(\P_{H}(E))$.
%\end{itemize}
%
A \emph{hypothetical expected utility} (HEU) maximizer is then given by  $\< \P, \mu \>$ where $\P$ is a coherent interpretation and $\mu$ is a measure defined over $\W$,
such that the following relation holds:\footnote{\label{ft:finite}Where $\triangle$ denotes the symmetric difference. This condition ensures that implication, as derived from $\pi$, can be faithfully recovered from probabilistic judgements. It is without loss of generality in finite state spaces, since, for any $\< \P, \mu \>$ there exists a $\< \P', \mu \>$ that satisfies the conditions and represents the same preference in the sense of \eqref{rep} via the map 
$$\P': H \mapsto \bigcap \{ \P(G) \mid \mu(\pi(H) \triangle \pi(G)) = 0\}.$$ Notice, also, that this is always always met when $\mu$ has full support over a finite state space.}
\begin{align}
\label{heucond}
\text{If} \quad \mu(\pi(H) \triangle \pi(G)) = 0 \quad \text{then}\quad \pi(H) = \pi(G).
\end{align} 
The value of the act $f$, is then given by
\begin{align}
\label{rep}
\tag{\textsc{heu}}
V(f) =  \int^\mathscr{C} f \ \textup{d}(\mu\circ \pi)
\end{align}
which for the simple case of bets, resolves to $V(b_H) =\mu(\P(H))$. As usual, we say that $\< \P, \mu \>$ \emph{represents} $\s$ if the corresponding value function orders acts exactly as $\s$ does.
%
%\begin{lemma}
%For all $H$ and $E, F \subseteq H$, (i) $\P_H(\emptyset) = \emptyset$, (ii) $\P_H(H) = H$ and (iii) $\P_H(E \cup F) = \P_H(E) \cup \P_H(F)$.
%\end{lemma}
%
%\begin{proof}
%Part (i) follows from
%\begin{align*}
%\P_H(\emptyset) = \P_H(E \cap E^c) &= \P_H(E) \cap \P_H(E^c) && \text{(by \ref{and})} \\
%&= [\P_H(E)]^c \cap \P_H(E) && \text{(by \ref{neg})} \\
%&= \emptyset.
%\end{align*}
%Part (ii) from part (i) to \ref{neg}. Part (iii) follows from DeMorgan's Laws. 
%\end{proof}
%
%\begin{lemma}
%For all $H$ and $E \subseteq H$, 
%$$E \setminus ker(\P_H) \subseteq \P_H(E) \subseteq E \cup ker(\P_H).$$
%\end{lemma}
%
%\begin{proof}
%Towards the left inclusion, let $\w \in E$ and $\w \notin \P_H(E)$. We will show $\P_H(\w) = \emptyset$.
%We have 
%\begin{align*}
%\P_H(\w) &= \P_H(E \cap \w) && \text{(since $\w \in E$)} \\
%&= \P_H(E) \cap \P_H(\w) && \text{(by \ref{and})} \\
%&= \P_H(\P_H(E)) \cap \P_H(\w) && \text{(by \ref{idp})} \\
%&= \P_H(\P_H(E) \cap \w) && \text{(by \ref{and})} \\
%&= \P_H(\emptyset) && \text{(since $\w \notin \P_H(E)$)} \\
%&= \emptyset.
%\end{align*}
%Establishing the right inclusion follows from an analogous sequence, taking some $\w \in \P_H(E)$ and $\w \notin E$ and showing $\P_H(\w) = \emptyset$.
%%\begin{align*}
%%\P_H(\w) &= \P_H(\P_H(E) \cap \w) && \text{(since $\w \in \P_H(E)$} \\
%%&= \P_H(\P_H(E)) \cap \P_H(\w) && \text{(by \ref{and})} \\
%%&= \P_H(E) \cap \P_H(\w) && \text{(by \ref{idp})} \\
%%&= \P_H(E \cap \w) && \text{(by \ref{and})} \\
%%&= \P_H(\emptyset) && \text{(since $\w \notin E$)} \\
%%&= \emptyset.
%%\end{align*}
%\end{proof}

\subsection{Implication from Preference} 
\label{sec:ifp}

The DM's perception of implication can be identified from her preferences over acts.
In particular, we say that the decision maker, with preference $\s$, \emph{reveals that she perceives that the hypothesis $H$ implies the hypothesis $G$} if $b_{G} \sim b_{H\cup G}$, and we parsimoniously right $H\im G$.\footnote{The notation here is supposed to be evocative, if a little on the nose, in that it makes an arrow with the preference symbol. This has the bennefit that it allows us to discuss the implication of two DMs, $\s_1$ and $\s_2$, using $\im_1$ and $\im_2$.}  In such a case, the DM is equally willing to bet on $G$ as on the disjunctive hypothesis that either $H$ \emph{or} $G$ is true. In general, winning a bet given either $H$ or $G$ provides more ways to win, and so the DM's indifference betrays a belief that every (payoff relevant) contingency not precluded by $H$ is likewise not precluded by $G$. In other words, the DM believes it impossible that $H$ is true and $G$ is not.
%For each hypothesis, $H$, let $\d[H] = \{G \mid b_{H\cap G} \sim b_H\}$ collect hypotheses the DM perceives as being implied by the hypothesis $H$. %If a given contingency is relevant when considering $H$---in the sense that it is payoff relevant in relation to $H$--- it is \emph{still} relevant when considering the joint hypothesis $G$ and $H$.
%Finally, define $H^\star = \bigcap \d[H]$ to be the joint hypothesis over all things perceived to be implied by $H$. 
%

\begin{proposition}
\label{prop:derivedfromrep}
Let $\<\pi,\mu\>$ represent $\s$. Then $\im$ is derived from $\pi$. 
\end{proposition}

If $\W$ was infinite, the 1-1 correspondence between betting behavior and subjective implication fails. This is because the DM, when considering her preference over bets, ignores 0-probability events, while implication is does not. According to betting behavior: $b_\emptyset \sim b_{\emptyset \cup H}$ for any $\mu$-probability 0 event $H$. But $\pi$ would then send $\emptyset$ to something containing the union of all zero probability events, which may itself be of positive probability. Thus, in uncountable state spaces, implication must be elicited in a more subtle way.

\section{Axioms}
\label{section:axioms}

\begin{ax}{A}{chq}{Choquet}
$\s$ is represented by a Choquet Expected Utility. 
\end{ax}

%\begin{ax}{A}{mod}{modularity} If $f \sim \min\{f,g\}$ then $\min\{f,h\} \sim \min\{f,g,h\}$
%\end{ax}

\begin{ax}{A}{mod}{modularity} If $g \sim g \lor f$ then $g \lor h \sim g \lor h \lor f$.
\end{ax}

It is worth briefly examining \rax{mod} in the context of individual bets:
$$b_{G} \sim b_{G \cup H} \text{ implies } b_{G \cup F} \sim b_{G \cup F \cup H},$$
From this it becomes clear that \rax{mod} is indeed the modularity axiom of \cite{kreps1979representation}.\footnote{Perhaps a mere curiosity: We can define a preference relation $\mathbf{R}$ over the set of non-empty hypotheses given by $H\mathbf{R}G$  if $b_{H} \s b_{G}$. Then, \rax{chq} implies \citeauthor{kreps1979representation}' preference for flexibility axiom, as well ensuring $\mathbf{R}$ is a weak order. Therefore, $\mathbf{R}$, which when derived from an HEU preference is represented by $\mu \circ \pi$, also has a Krepsian representation in that it can be represented by a subjective state space and set of state dependent utilities.} There it lends an entirely different interpretation but plays a technical similar role.

%Given a hypothesis $H$ call a contingency $\w \in H$, \emph{perceived relevant to the hypothesis $H$} if $b_H \succ b_{H\setminus \w}$ for some $b \in \R_+$. 
%
%\begin{ax}{A}{rel}{relevance}
% $b_{H} \succ b_{H \setminus \w}$ and $b_{H'} \succ b_{H' \setminus \w}$ implies $b_{H\cap H'} \succ b_{(H\cap H') \setminus \w}$
%\end{ax}

%\begin{ax}{A}{rel}{relevance}
%If $F \in \d[H\cap H']$, then there exists a $G \in \d[H]$ and a $G' \in \d[H']$ such that $F = G\cap G'$.
%\end{ax}

\begin{ax}{A}{rel}{relevance}
If $h\lor h' \sim  h\lor h' \lor f$, then there exists a $g,g' \in \F$ such that $h \sim h \lor g$, $h' \sim h' \lor g'$ and $f = g \lor g'$.
\end{ax}

\rax{rel} states that the DM perceives $F$ to be implied by the joint hypothesis of $H$ and $H'$, then $F$ can be decomposed into two weaker hypotheses: one, $G$, that follows from $H$ and the other $G'$ that follows from $H'$. \rax{rel} is related, from a different vantage, to a notion of dynamic consistency: first learning $H$ then learning $H'$ yields the same implications as learning the two simultaneously.  

\begin{theorem}
\label{thm:dom}
If $\s$ satisfies \textsc{\textbf{A4}} and \textsc{\textbf{A5}} then \textup{$\im$} is derived from a weakly coherent interpretation. If in addition, $\s$ satisfies \textsc{\textbf{A7}} then \textup{$\im$} is derived from a coherent interpretation. 
\end{theorem}

Theorem \ref{thm:dom} provides the existence of interpretation, needed for a HEU representation. What remains, then, is to ensure the existence of a probability over the state space giving rise to the DM's preference when combined with this projection. Normally, the existence of such a measure would be guaranteed by a straightforward independence or linearity restriction. This approach, unfortunately, does not apply so straightforwardly here. The impediment lies in the fact that the DM's preference only measures specific subsets of the state space (sets in the image of $\P$). This collection might be very sparse (i.e, far from being an algebra), and so, we need stronger conditions on our observable to ensure the local linearity can be lifted to the entire state-space. 

This restriction comes in the form of \emph{total monotonicity}, a property that plays a key (albeit very different) roll in the theory of belief functions, following \cite{dempster1967upper} and \cite{shafer1976mathematical}. Towards introducing this within our framework, we need the following bits of notation. For each $n$, let $\mathcal I(n)$ collect all non-empty subsets of $\{1\ldots n\}$. Further, for an indexed set of $n$ hypotheses, $\{H_i\}_{i\leq n}$, set for each $I \in \mathcal I(n)$ some hypothesis $H_I$ such that $\d[H_I, \im] = \bigcap_{i \in I} \d[H_i,\im]$.

\begin{remark}
If $\s$ satisfies \rax{chq} and \rax{mod} then it is the conclusion of Theorem \ref{thm:dom} and Lemmas \ref{lem:pi} and \ref{lem:charu} that for any $\{H_i\}_{i\leq n}$ and $I \in \mathcal I(n)$, such a hypothesis $H_I$ exists. Moreover, if $\pi$ is the unique weakly coherent interpretation from which $\im$ is derived, then $\pi(H_I) = \bigcap_{i \in I} \pi(H_i)$.
\end{remark}

The hypotheses that imply $H_I$ are exactly those that imply each $H_i$. Thus, a coherent DM interprets $H_I$ as the intersection of her interpretation of the $H_i$'s. While $H_I$ need not be unique, this multiplicity is non-threatening since the DM's betting behavior is, by definition, invariant across this class.

%\begin{ax}{A}{i/e}{Inclusion / Exclusion}
%Let 
%\begin{equation}
%\label{lin1}
%\d[G, \im] \subseteq \bigcap_{i \leq n} \d[H_i, \im].
%\end{equation} 
%Then for any hypothesis $F$,
%\begin{equation}
%\label{lin2}
%b_{H_I}\sim \alpha^I b_F  \text{ for all } I \in \mathcal I(n) 
%\quad \text{implies} \quad b_G \s (\sum_{\mathcal I(n)} (-1)^{|I|+1} \alpha^I)b_F
%\end{equation}
%where $\alpha^I \in \R_+$ for each $I \in \mathcal I(n)$, and where the consequent of \eqref{lin2} holds with indifference when \eqref{lin1} holds with equality.
%\end{ax}

\begin{ax}{A}{i/e}{Inclusion / Exclusion} Let $G = \bigcup_{i \leq n} H$. Then for any $F$,
\begin{equation*}
%\label{lin2}
b_{H_I}\sim \alpha^I b_F  \text{ for all } I \in \mathcal I(n) 
\quad \text{implies} \quad b_G \sim (\sum_{\mathcal I(n)} (-1)^{|I|+1} \alpha^I)b_F
\end{equation*}
where $\alpha^I \in \R_+$ for each $I \in \mathcal I(n)$.
\end{ax}

To see what \rax{i/e} entails, consider the case where $G = H_1 \cup H_2$. By \r{d}, we have $\pi(G) = \pi(H) \cup \pi(H')$. So then, since $H_{\{1\}} = H_1$ and $H_{\{2\}} = H_2$, \rax{i/e} states that betting on $G$ valued equally to the sum of the values of betting on $H_1$ and $H_2$:

 if $\alpha_1 b_F \sim b_{H_1}$ and $\alpha_2 b_F \sim b_{H_2}$ then $(\alpha_1 + \alpha_2) b_F \sim b_{G}$. Of course, when $H_1$ and $H_2$ are not disjoint, a bet on $H_1$ and a bet of $H_2$ includes getting paid twice in the event they both are true. To correct for this, \rax{i/e} requires that we subtract this additional term: if $\alpha_1 b_F \sim b_{H_1}$ and $\alpha_2 b_F \sim b_{H_2}$ and $\alpha_{\{1,2\}} b_F \sim b_{H_1 \cap H_2}$ then $(\alpha_1 + \alpha_2 - \alpha_{\{1,2\}}) b_F \sim b_{G}$. Then, \rax{i/e} requires that this pattern holds for larger collections of hypotheses.

\begin{theorem}
\label{thm:rep}
$\s$ satisfies \textsc{\textbf{A5-8}} if and only if it has a Hypothetical Expected Utility representation, $\<\mu,\pi\>$, Moreover, $\pi$ is unique and $\mu$ is uniquely defined over the algebra generated by the image of $\pi$. 
\end{theorem}

\section{Comparative Hypothetical Reasoning}
\label{sec:cs}

Where there are multiple HEU DMs facing the same uncertain environment, there is a natural ordering on their hypothetical reasoning. A DM is a better hypothetical reasoner than another DM whenever her interpretations are always closer to the truth. To make this precise let $\s_1$ and $\s_2$ be two HEU preferences defined on a common state-space, and let $\<\pi_1,\mu_1\>$ represent $\s_1$ and $\<\pi_2,\mu_2\>$ represent $\s_2$. Let $\im_1$ and $\im_2$ denote the respective derived subjective implication relations. 

Formally, say $\s_1$ is \emph{a better hypothetical reasoner} than $\s_2$ if $H \subseteq \pi_1(H) \subseteq \pi_2(H)$ for all hypotheses $H$. Not all DMs can be compared in the manner, since it is possible to have a more accurate interpretation for some hypotheses and a less accurate one for others. The `better hypothetical reasoner' relation is a partial order over all possible interpretations with $\pi(H) = H$ being the maximal element and $\pi(H) = \W$ being the minimal element. A desirable property of this definition is that it in no way depends on $\mu$ and so does not require the DMs to entertain the same probabilistic judgements. 

The characterization of the better hypothetical reasoner is simple given the machinery developed above:

\begin{proposition}
\label{prop:cs}
Let $\s_1$ and $\s_2$ be two HEU preferences defined on a common state-space. Then the following are equivalent: 
\begin{enumerate}
\item $\s_1$ is a better hypothetical reasoner than $\s_2$, and
\item $G \im_1 H$ implies $G \im_2 H$ for all hypotheses $H$ and $G$.
\end{enumerate}
\end{proposition}

Proposition \ref{prop:cs} shows that DM 1 is a better reasoner than DM 2 exactly when she perceives less implications. So, this result reinforces the conception of hypothetical reasoning as the ability to properly understand implication.

\section{Ambiguity Attitudes and Dual Models}
\label{sec:aa}

A DM's attitude towards subjective uncertainty has long been connected to her preference for hedging: her desire to smooth consumption by hedging between two uncertain acts. Of course, a (strict) preference or dis-preference for hedging is a violation of subjective expected utility.

\begin{ax}{A}{al}{Aversion to Hedging}
If $f \s g$, then $f \s \frac12f + \frac12g$.
\end{ax}

\addtocounter{ax}{-1}
\begin{ax'}{A}{aa}{Preference for Hedging}
If $f \s g$, then $\frac12f + \frac12g \s g$.
\end{ax'}

Call a DM \emph{ambiguity loving} if she displays an aversion to hedging and \emph{ambiguity averse} if she displays the 
The HEU decision maker, by virtue of misinterpreting events, displays an aversion to hedging. 

\begin{corollary}
\label{cor:aa}
If $\s$ has a Hypothetical Expected Utility representation then it satisfies \rax{al}.
\end{corollary}

It is well known, e.g., see \cite{schmeidler1989subjective}, that a Choquet EU decision maker is ambiguity loving if and only if the representing capacity is concave: that is if $\nu(G\cap H) + \nu(H \cup G) \leq \nu(H) + \nu(G)$. When $\nu = \mu\circ\pi$ for a measure $\mu$ and a coherent interpretation $\pi$ this always holds, as can be concluded by the observation that that $\pi(H \cap G) \subseteq \pi(H) \cap \pi(G)$ and $\pi(G\cup H)=\pi(H) \cup \pi(G)$. 

Often, however, it is not ambiguity loving, but ambiguity \emph{averse} behavior that economists want to explain. Towards this, it is worth possibly considering the dual model of the one presented here. Call an interpretation $\pi$ dual-coherent if it satisfies along with \r{m} and \r{i}:
\begin{itemize}[leftmargin=20ex]
\item[\mylabel{t'}{T'}{Truth'}] $\P(H) \subseteq H$, and
\item[\mylabel{c'}{C'}{ Consistency'}] $\P(H) \cap \P(G) \subseteq \P(H\cap G)$.
\end{itemize}
Axiomatizing a dual-coherent interpretation, either from an implication relation or a preference relation, is a straightforward exercise in exchanging objects for their dual counterparts. 

Call a DM a dual-HEU maximizer if her preferences are represented by some $\<\mu,\pi\>$ with a dual-coherent $\pi$. It should be clear that such a DM displays a preference for hedging and is therefore ambiguity averse. More interestingly, is, perhaps, the interpretation of ambiguity aversion that is afforded by such a model. Here the DM \emph{fails to perceive all contingencies consistent with a hypothesis} as evidenced by \r{t'}. Such a DM find value in hedging because she cannot fail to recognize a contingency is part of with $H$ or $G$ but recognize that it is contained by $H \cup G$.

\section{Discussion and Related Literature}
\label{sec:lit}

In contrast to many biases in decision making (uncertainty aversion, improper Bayesian updating, reference dependence, etc) there are few, if any, simple yet general models of flawed hypothetical reasoning. 
Failures in hypothetical thinking have been oft cited in psychology and economics as a reason for empirically observed deviations form normative predictions. In a forceful exhibit of illogicality, \cite{tversky1983extensional} found evidence of what they called the \emph{conjunction fallacy}, wherein subjects rank the likelihood of statement of the form `$p \textsc{ and } q$' as strictly higher than the statement `$p$'. Since this is a logical impossibility irrespective of the interpretation of propositions `$p$' and `$q$,' their findings provide unambiguous evidence imperfect hypothetical reasoning. Such behavior cannot be rationalized by the present model as it violates monotonicity via \r{m}. 

More recently, there have been experimental economic studies where subjects take dominated (or inconsistent) actions in apparent violation of cogent hypothetical judgments. \cite{esponda2014hypothetical}, in a very clean design, find that subjects in a voting game play dominated actions by not conditioning on being pivotal. \cite{martinez2019failures} examine such failure of contingent reasoning in the presence of uncertainty and find the effect exacerbated. \cite{agranov2020non} report subject's deliberate randomization across identical decision problems in a way inconsistent with maximizing payoff with respect to any known model of beliefs. 

There is a small literature that discusses relation between contingent and probabilistic reasoning. In a similar spirt to this paper, \cite{mukerji1997understanding} considers a DM who maps an objective state space (i.e., the one that governs payoffs) into a subjective one. The author then shows that if the inverse mapping fails to preserve unions of events the DM will display ambiguity aversion. A critical difference is between \cite{mukerji1997understanding} and the incumbent paper, is that it takes as given the DM's mapping between the objective and subjective state spaces, rather than identifying it from behavior (i.e., preferences over acts) or a subjective implication relation. 

\cite{esponda2019contingent} show that many of the experimental results that have been attributed to a failure of contingent reasoning can be recast as a violation of Savage's sure-thing-principle. The present paper furthers this general program but providing a particular definition of contingent (i.e., hypothetical) in terms patterns of choice where are independent fo the particular choice environment and showing that such preference necessarily violate the sure-thing-principle (or, more accurately, its analogue for convex spaces, the independence axiom). 

\cite{piermont20} also derives an \emph{implication relation} from preferences, albeit preferences over a more abstract space of propositions. There the main focus is on non-monotone implications relations, in other words, that violate \r{m}. In the present formulation, precisely because of \r{m}, the DM perceives too many implications. Relaxing \r{m} allows for a DM whose hypothetical judgments fail in the opposite direction, whereby she perceives too few implications. Such non-monotone implication relations can explain extreme (although not uncommon) violations of logical rationality such as the \emph{conjunction fallacy} as introduced by \cite{tversky1983extensional}. In \cite{piermont20} the state-space is a derived object but it is not unique. 

This paper also draws heavily on the theory of \emph{closure operators} from order theory and topology \cite{kuratowski2014topology}. Although, to the best of my knowledge, the main results are all novel, the objects of study are far from. Indeed, many of the definitions and some of the interim results (lemmas) can be found scattered through this literature. A coherent interpretation, $\pi$, that also satisfies the stipulation that $\pi(\emptyset) = \emptyset$ is known as a (topological) closure operator and it is well known that the set $\Pi = \{H \mid \pi(H) = H\}$ forms a topology (and for any topology on $\W$, the map $H \mapsto \textup{cl}(H)$ is a closure operator (hence a coherent interpretation)). Likewise, its dual (that preserves $\W$) is known as an interior operator. This should come as no surprise to those initiated to the field of modal logic, where the connection between topology and logic of knowledge and belief goes back to as least \cite{mckinsey1941solution}; see \cite{parikh2007topology} for a survey.

\appendix 

\section{Proofs}
\label{sec:proofs}

%In this section, we adopt the notational convention of associating each hypothesis $H$ with $1_H$. Thus $H \s G$ is written to mean $1_H \s 1_G$.
%
%\bigskip 

\begin{pproof}{prop:drvw}
\textbf{Necessity.} Necessity of \rax{trv} and \rax{mon} are obvious. We will show the necessity of \rax{ded}. 
Let $H_i \implies G$ for all $i$ in some index set $\mathcal I$. Then $\P(H_i) \subseteq \P(G)$---and therefore by \r{t}, $H_i \subseteq \P(H_i) \subseteq \P(G)$---for each $i \in \mathcal{I}$. So $\bigcup_{i \in \mathcal I} H_i \subseteq \P(G)$. By \r{m}, this implies $\P\big(\bigcup_{i \in \mathcal I} H_i \big) \subseteq \P(\P(G))$. Lastly, \r{i} yeilds $\P\big(\bigcup_{i \in \mathcal I} H_i \big) \subseteq \P(\P(G))$ which shows that $\bigcup_{i \in \mathcal I} H_i \implies G$.

\textbf{Sufficiency.}  Define $\P$ as $H \mapsto \bigcup \d[H]$. We first show that $\implies$ is derived from $\P$. Let $G \implies H$. Then by \rax{trv}, $\d[G] \subseteq \d[H]$ and hence $\pi(G)\subseteq \pi(H)$. Now let $\pi(G)\subseteq \pi(H)$. By \rax{ded}, $\pi(H) \implies H$. By \rax{mon}, $G \in \d[G]$ and hence $G \subseteq \pi(G)$. Therefore, $H \implies \pi(H) \subseteq \pi(G) \subseteq G$ and so by \rax{mon} and \rax{trv}, $H \implies G$.

Next we verify that $\P$ is coherent. By \rax{mon} $G \in \d[H]$ for all $G \subseteq H$. First notice this implies $H \in \d[H]$, or that, $H \subseteq \pi(H)$: we have \r{t}. Second notice that time implies $\d[G] \subseteq \d[H]$ for $G \subseteq H$: we have \r{m}. Finally, by \rax{ded}, $\pi(H) \implies H$, by \rax{mon} $H \implies \pi(H)$. \rax{trv} therefore delivers $\d[H] = \d[\pi(H)]$ so $\pi(H) = \pi(\pi(H))$: we have \r{i}.

\textbf{Uniqueness.} Given $\implies$, let $\iff$ be the symmetric component. It follows from \rax{mon} and \rax{trv} that $\iff$ is an equivalence relation: let $[H]$ denote the equivalence classes. From \rax{ded} we have that $\bigcup [H] \in [H]$. Now, notice, that for any $\pi$ such that $\implies$ is derived from $\pi$, it must be that $\pi(\bigcup [H]) = \bigcup [H]$. Indeed, by \r{i}, we have $\pi(\pi(\bigcup [H])) = \pi(\bigcup [H])$ so $ \pi(\bigcup [H]) \iff \bigcup [H]$ or that $\pi(\bigcup [H]) \in [H]$. Thus, $\pi(\bigcup [H]) \subseteq \bigcup [H]$ and other inclusion follows from \r{t}.
Finally, notice that for all $G \in [H]$, $\pi(G)$ must equal $\pi(H)$ and so $\pi$ is uniquely determined by its value, $\pi(\bigcup [H])$, for each equivalence class. 
\end{pproof}

\begin{pproof}{prop:drv}
\textbf{Necessity.} In light of Proposition \ref{prop:drvw} only the necessity of \rax{dcmp} remains. To see \rax{dcmp}, notice that if  $F \implies H \cup H'$ then $F \cap \pi(H) =^{df} G \implies H$. Indeed, 
$G = F \cap \pi(H) \subseteq \pi(H)$. 
%what the fuck
%\begin{align*}
%G &= F \cap (\pi(H )\setminus \pi(H')) &&  \\
%& \subseteq \pi(F)  \cap (\pi(H )\setminus \pi(H')) && \text{(by \r{t})}  \\
%& \subseteq \pi(H \cup H')  \cap (\pi(H )\setminus \pi(H')) && \text{(since $F \implies H \cup H' $)} \\
%& \subseteq (\pi(H) \cup \pi(H'))  \cap (\pi(H )\setminus \pi(H')) && \text{(by \r{c})} &&  \\
%&= \pi(H)
%\end{align*}
So, using properties \r{m} and \r{i}, we obtain $\pi(G) \subseteq \pi(\pi(H)) = \pi(H)$. Moreover, since $F \implies H\cup H'$, \r{t} and \r{c} insinuate that $F \subseteq \pi(F) \subseteq \pi(H \cup H') \subseteq \pi(H) \cup \pi(H')$. Thus, (defining $G'$ in analog to $G$ above) we have $F = G \cup G'$.

\textbf{Sufficiency.} 
We will show that the (weakly coherent) $\P: H \mapsto \bigcup \d[H]$ satisfies \r{d} when $\implies$ satisfies \rax{dcmp}. 
%Now, by \rax{mon} and \rax{trv} we have $\d[H] \cup \d[H'] \subseteq \d[H\cup H']$. So $\pi(H)\cup \pi(H) \subseteq \pi(H\cup H')$. 
Fix $H$ and $H'$. By \r{i} it must be that  $\pi(\pi(H\cup H'))=\pi(H\cup H')$: so we have $\pi(H \cup H') \implies H \cup H'$. Applying \rax{dcmp} yields a $G, G'$, such that $G \in \d[H]$, $G' \in \d[H']$ and $\pi(H\cup H') = G \cup G'$.
Thus $\pi(H\cup H') = G\cup G' \subseteq \pi(H) \cup \pi(H')$.
\end{pproof}

\begin{pproof}{prop:derivedfromrep}
Let $\pi(H) \subseteq \pi(G)$. Then $\pi(H\cup G) = \pi(H)\cup \pi(G) = \pi(G)$, and so $V(b_H \lor b_G) = V(b_{H \lor G}) = \mu(\pi(H \cup G)) = \mu(\pi(G)) = V(b_G)$. Now let $V(b_H \lor b_G) = V(b_G)$ so that $\mu(\pi(H \cup G)) = \mu(\pi(G))$. \r{m} implies $\pi(H) \subseteq \pi(H\cup G)$, so we have $\mu(\pi(H\cup G) \triangle \pi(G))=0$, implying $\pi(H\cup G) = \pi(G)$, or $\pi(H) \subseteq \pi(H\cup G) = \pi(G)$.
\end{pproof}

\begin{tproof}{thm:dom}
\begin{enumerate}
\item[\rax{trv}:] Let $H \im G$ and $G \im F$. Then $b_G \sim b_G \lor b_H$ and $b_F \sim b_F \lor b_G$. By \rax{mod}, these relations yield $b_G \lor b_F \sim b_G \lor b_F \lor b_H$ and $b_F \lor b_H \sim b_F\lor b_H\lor b_G$. By \rax{chq}, $\s$ is transitive, liberal use of which indicates, $b_F \sim b_H \lor b_F$: hence $H \im F$.

\item[\rax{mon}:] Immediate from the definition of $\im$ and the reflexivity of $\s$ (as ensured by \rax{chq}).

\item[\rax{ded}:] Let $H \im G$ and $H' \im G$. Then $b_G \sim b_G \lor b_H$ and $b_G \sim b_G \lor b_{H'}$. \rax{mod}, and the latter relation yields $b_G \lor b_H \sim b_G \lor b_H \lor b_{H'}$. So, the transitivity of $\s$ produces $b_G \sim b_G \lor b_{H'} \lor b_{H} = b_G \lor b_{H\cup H'}$, and so $H \cup H' \im G$. Since we are working with a finite state-space this suffices.

\item[\rax{dcmp}:] Let $F \im H \cup H'$. Then $ b_H\lor b_{H'}  \sim b_H \lor b_{H'}\lor b_{F} $. \rax{rel}, and the latter relation yields the existence of some $g,g' \in \F$ such that $b_H \sim b_H\lor g$, $b_{H'} \sim b_{H'}\lor g'$ and $b_F = g \lor g'$. Set $G= \{\w \mid g(\w) > 0\}$ and $G'= \{\w \mid g'(\w) > 0\}$. Then it is a straightforward consequence of the Choquet representation, \rax{chq} that $G \im H$, $H' \im H'$ and $F = G \cup G'$.
\end{enumerate}
\end{tproof}

%\begin{lemma}
%\label{lem:charstar}
%$\w \in H\x$ if and only if $H \succ H\setminus \w$.
%\end{lemma}
%
%\begin{lproof}{lem:charstar}
%Assume that $\w \in H\x$. Then for all $F \in \d[H]$, $\w \in F$, and hence $ \W \setminus \w \notin \d[H]$. By the definition of $\d$ we have $H \not\sim H \cap (\W\setminus\w) = H \setminus \w$. By completeness and \rax{mon}, therefore, $H \succ H \setminus \w$.
%
%Now let $H \succ H\setminus \w$. Let $F \in \d[H]$. If $\w \notin F$, then by Lemma \ref{lem:dom}.\ref{upset}, $\W \setminus \w \in \d[H]$, which we assume is not the case: therefore, $\w \in F$ for all $F \in \d[H]$, and hence $H\x$.
%\end{lproof}

%\begin{lemma}
%\label{lem:piprop}
%An interpretation $\pi$ satisfies \r{m} and \r{c} if and only if it satisfies:
%\textup{
%\begin{itemize}[leftmargin=20ex]
%\item[\mylabel{d}{D}{ Distribution}] $\P(H\cup G) = \P(H) \cup \P(G)$.
%\end{itemize} }
%\end{lemma}
%
%
%
%\begin{lproof}{lem:pi}
%If: \r{c} is obvious. For $H \subseteq G$ we have $\P(H) \subseteq \P(H)\cup \P(G) = \P(H\cup G) = \P(G)$, so \r{m}. Only if: by \r{m} $\P(H) \subseteq \P(H\cup G)$ so $\P(H) \cup \P(G) \subseteq \P(H \cup G)$. The other inclusion is \r{c}.
%\end{lproof}

\begin{lemma}
\label{lem:pi}
Let $\pi$ be a weakly coherent. Then $\Pi = \{\pi(H) \mid H \subseteq \W\}$ is closed under arbitrary intersections. 
\end{lemma}

\begin{lproof}{lem:pi}
Let $H_i \in \Pi$ for all $i$ is some index set $\mathcal I$. Set $ H = \bigcap_{i \in \mathcal I} H_i$; we will show that $\pi(H) = H$, indicating that $H \in \Pi$. For each $i \in \mathcal I$ we have $H \subseteq H_i$ or by \r{m} and \r{i}, $\pi(H) \subseteq \pi(H_i) = H_i$. Hence $\pi(H) \subseteq  \bigcap_{i \in \mathcal I} H_i = H$. The other direction follows directly from \r{t}.
\end{lproof}

\begin{lemma}
\label{lem:charstar}
Let $\implies$ be derived from $\pi$ with $\pi$ weakly coherent. Then $H \implies G$ if and only if $H \subseteq \pi(G)$.
\end{lemma}

\begin{lproof}{lem:charstar}
Only If: Since $\implies$ is derived from $\pi$, $H \implies G$ is equivalent to $\pi(H) \subseteq \pi(G)$. By \r{t}, this implies $H \subseteq \pi(H) \subseteq \pi(G)$. If: $H \subseteq \pi(G)$ implies by \r{m} that $\pi(H) \subseteq \pi(\pi(G))$ and hence by \r{i}, $\pi(H) \subseteq \pi(G)$.
\end{lproof}

\begin{lemma}
\label{lem:charu} Let $\implies$ be derived from $\pi$, with $\pi$ weakly coherent.
Then: $\pi(G) = \bigcap_{i \leq n}\pi(H_i)$ if and only if $\d[G] = \bigcap_{i \leq n} \d[H_i]$.
% and the consequent holds with equality if and only if the antecedent does. 
\end{lemma}

\begin{lproof}{lem:charu}
To the the `if' direction assume $\d[G] = \bigcap_{i \leq n} \d[H_i]$. Then, for any $F \subseteq \W$, we have
\begin{align*}
F \subseteq \pi(G) &\iff F \in \d[G] && \text{(by Lemma \ref{lem:charstar})} \\
&\iff F \in  \bigcap_{i \leq n} \d[H_i] && \text{(by the assumption)}  \\
&\iff F \subseteq \pi(H_i) \text{ for all } i \leq n && \text{(by Lemma \ref{lem:charstar})} \\
&\iff F \subseteq \bigcap_{i \leq n}\pi(H_i).
\end{align*}
%Notice, that in addition if $\d[G] = \bigcap_{i \leq n} \d[H_i]$, the all implications are bi-directional.
The only if direction is a rearrangement of the same argument.
\end{lproof}

\begin{tproof}{thm:rep}
\textbf{Necessity.} Let $\< \P, \mu \>$ be an HEU  representation of $\s$. 

\rax{chq}. This is immediate. 

\rax{mod}. Fix some $f,g,h \in \F$ such that $g \sim g \lor f$. Set $x_1 \ldots x_n$ to be the distinct values taken by $f$, $g$, and $h$ in increasing order (i.e., $x_i \leq x_{i+1}$) set $x_0 = 0$. For an act $f'$ let $F'_{x_i} = \{\w \mid f'(\w) \geq x_i\}$ (i.e., for each act, the corresponding capital letter is the hypothesis). Notice also, for any acts $f',f''$, we have $F_{x_i} \cup F'_{x_i} = \{\w \mid (f' \lor f'')(\w) \geq x_i\}$.

Then by the representation we have
\begin{equation}
\label{eq:chqeq1}
 \sum_{i=1}^{n} (x^i - x^{i-1})\mu(\pi(G_{x_i})) =  \sum_{i=1}^{n} (x^i - x^{i-1})\mu(\pi(G_{x_i} \cup F_{x_i}))
\end{equation}
Since $G_{x_i} \subseteq G_{x_i} \cup F_{x_i}$ for all $i$, to maintain equality of \eqref{eq:chqeq1} it must be that $\mu(\pi(G_{x_i})) =  \mu(\pi(G_{x_i} \cup F_{x_i}))$ for all $i$---so we have $\mu(\pi(G_{x_i}) \triangle \pi(G_{x_i} \cup F_{x_i}))=0$, implying by \eqref{heucond} that $\pi(G_{x_i}) = \pi(G_{x_i} \cup F_{x_i})$.
Finally, since $\pi$ is coherent, and therefore satisfies \r{d}, we have $\mu(\pi(G_{x_i} \cup H_{x_i})) = \mu(\pi(G_{x_i} \cup H_{x_i}\cup F_{x_i}))$, which by the representation again, indicates that $g\lor h \sim g \lor h \lor f$. $\s$ satisfies \rax{mod}. 

\rax{rel}. Fix some $f,h',h \in \F$ such that $h \lor h' \sim  h \lor h' \lor f$. Set $x_1 \ldots x_n$ to be the distinct values taken by $f$, $h$, and $h'$ in increasing order (i.e., $x_{x_i} \leq x_{i+1}$) set $x_0 = 0$. Using the notation above, the representation provides that $\mu(\pi(H_{x_i} \cup H'_{x_i})) = \mu(\pi(H_{x_i} \cup H'_{x_i} \cup F_{x_i}))$ for all $i$; this implies via \eqref{heucond} that $\pi(H_{x_i} \cup H'_{x_i}) = \pi(H_{x_i} \cup H'_{x_i} \cup F_{x_i}) = \pi(H_{x_i} \cup H'_{x_i}) \cup \pi(F_{x_i})$. In particular this implies $F_{x_i} \subseteq \pi(H_{x_i} )\cup\pi( H'_{x_i})$.

Now define $g,g' \in \F$ as
%\begin{align*}
%g(\w) &= \begin{cases}
%\min\{ f(\w), h(\w)\} &\text{ if } \w \in (\pi(H'_{h(\w)}) \setminus  \pi(H_{h(\w)})) \\
%f(\w) &\text{ otherwise }
%\end{cases}
% \\
%g'(\w) &= \begin{cases}
%\min\{ f(\w), h'(\w)\} &\text{ if } \w \in (\pi(H_{h(\w)}) \setminus  \pi(H'_{h(\w)})) \\
%f(\w) &\text{ otherwise }
%\end{cases}
%\end{align*}
\begin{align*}
g(\w) &= \begin{cases}
\min\{ f(\w), h(\w)\} &\text{ if } \w \in \W \setminus \pi(H_{h(\w)}) \\
f(\w) &\text{ otherwise }
\end{cases}
 \\
g'(\w) &= \begin{cases}
\min\{ f(\w), h'(\w)\} &\text{ if } \w \in \W \setminus \pi(H'_{h'(\w)}) \\
f(\w) &\text{ otherwise }
\end{cases}
\end{align*}
It is immediate that $f = g \lor g'$.

Moreover, for each $i$ we have: $G_{x_i} = F_{x_i} \cap \pi(H_{x_i} )$. Thus, replicating the Necessity proof of Proposition \ref{prop:drvw}, we obtain $\pi(G_{x_i} ) \subseteq \pi(H_{x_i})$ for all $i$, and therefore, that $h \sim h \lor g$.

%Moreover, for each $i$ we have (letting $S = \supp(\mu)$),
%\begin{align*}
%G_{x_i} &= F_{x_i} \cup (\pi(H_{x_i} )\setminus \pi(H'_{x_i})) &&  \\
%& \supseteq \pi(F_{x_i})  \cup (\pi(H_{x_i} )\setminus \pi(H'_{x_i})) && \text{(by \r{t})}  \\
%& \supseteq (\pi(H_{x_i} \cap H'_{x_i}) \cap S)  \cup (\pi(H_{x_i} )\setminus \pi(H'_{x_i})) && \text{(by \eqref{eq:hhf})} \\
%& \supseteq (\pi(H_{x_i} \cap H'_{x_i}) \cup (\pi(H_{x_i} )\setminus \pi(H'_{x_i}))) \cap S &&  \\
%& = ((\pi(H_{x_i}) \cap \pi(H'_{x_i})) \cup (\pi(H_{x_i} )\setminus \pi(H'_{x_i}))) \cap S &&   \text{(by \r{d}} \\
%& = \pi(H_{x_i}) \cap S && 
%\end{align*}
%and so $G_{x_i}^c \subseteq \pi(H_{x_i})^c \cup S^c$. Now consider the act $\tilde h = h - h\land g$. We have
%\begin{align*}
%\pi(\tilde H_{x_i}) &=  \pi(H_{x_i} \cap G^c_{x_i})&&  \\
%&=  \pi(H_{x_i}) \cap \pi(G^c_{x_i})&& \text{(by \r{d} )}   \\
%&\subseteq \pi(H_{x_i}) \cap \pi (\pi(H_{x_i})^c \cup S^c)  && \text{(by \r{m} and the above identity)}  \\
%&= \pi(\pi(H_{x_i})) \cap \pi (\pi(H_{x_i})^c \cup S^c)  && \text{(by \r{i}})  \\
%&= \pi(\pi(H_{x_i}) \cap (\pi(H_{x_i})^c \cup S^c))  && \text{(by \r{d}})  \\
%&= \pi(\pi(H_{x_i})  \cap S^c)  &&  \\
%&\subseteq \pi(S^c)  &&  \\
%&\subseteq S^c  && \text{(by \r{t}})
%\end{align*}
%So, $\mu(\pi(\tilde H_{x_i})) = 0$ for all $i$ and therefore $h-h\land g \sim 0$. But $h$ and $h\land g$ are co-monotone, so by the Choquet representation $h \sim h\land g$, as desired. The same argument, of course, shows that  $h' \sim h'\land g$.

\rax{i/e}. Fix some hypotheses $G,H_1 \ldots H_n, F$ and $\alpha^I \in \R_+$ for $I \in \mathcal I(n)$ such that 
$G = \bigcup_{i \leq n} H_i$ and  $b_{H_I}\sim \alpha^I b_F$ for all $I \in \mathcal I(n)$. By \r{d}, $\pi(G) = \bigcup_{i \leq n}\pi(H_i)$ and by the representation, and Lemma \ref{lem:charu}, we have $\alpha^I\mu(\pi(F)) = \mu(\pi(H_I)) = \mu(\cap_I \pi(H_i))$.

Since $\mu$ is a measure, hence totally monotone, we have
$$\mu(\pi(G))  = \sum_{\mathcal I(n)} (-1)^{|I|+1} \mu(\cap_I \pi(H_i)) = \sum_{\mathcal I(n)} (-1)^{|I|+1} \alpha^I\mu(\pi(F)), $$
which by the the representation yields  $ b_G\s (\sum_{\mathcal I(n)} (-1)^{|I|+1} \alpha^I)b_F$.

\textbf{Sufficiency.} Let $\s$ satisfy  \rax{chq}, \rax{mod}, \rax{rel} and \rax{i/e}. Let $\nu: 2^\W \to [0,1]$ be the capacity the represents 
$\s$ via a Choquet representation as dictated by \rax{chq}.
By Theorem \ref{thm:dom}, $\im$ satisfies \rax{trv}, \rax{ded}, \rax{mon}, and \rax{dcmp}. By Proposition \ref{prop:drv}, $\im$ is derived from a coherent interpretation $\pi: 2^\W \to 2^\W$. Let $\Pi = \{\pi(H) \mid H \subseteq \W\}$. By Lemma \ref{lem:pi}, $\Pi$ is a $\pi$-system (i.e., closed under intersections).

Define $\mu: \Pi \to [0,1]$ as $\mu: P \mapsto \nu(\pi^{-1}(P))$. $\mu$ is well defined: indeed, let $\pi(G) = \pi(H)$. Then $G \im H$ and $H \im G$ or, in other symbols, $b_H \lor b_G \sim b_H$ and $b_H \lor b_G \sim b_G$. Thus by \rax{chq}, $\nu(H) = \nu(G)$. Moreover, $\mu$ is grounded: $\mu(\emptyset)=0$: Let $H = \pi^{-1}(\emptyset)$ then $H \implies \emptyset$ or $b_H = b_H \lor b_\emptyset \sim b_\emptyset = 0$.

 Claim: For $P_1 \ldots P_n, Q \in \Pi$, we have the following:
	\begin{enumerate}
	\item If $\bigcup_{i\leq n} P_i = Q$ then $\sum_{\mathcal I(n)} (-1)^{|I|+1} \mu(\cap_I P_i) = \mu(Q)$, and
	\item If $\bigcup_{i\leq n} P_i \subseteq Q$ then $\sum_{\mathcal I(n)} (-1)^{|I|+1} \mu(\cap_I P_i) \leq \mu(Q)$.
	\end{enumerate}
Assume $\bigcup_{i\leq n} P_i = Q$. 
Let $H_i \in \pi^{-1}(P_i)$ and $G \in \pi^{-1}(Q)$. 
%Then by Lemma \ref{lem:charu}: $\d[G, \im] \subseteq \bigcap_{i \leq n} \d[H_i, \im]$. 
Since $\pi$ satisfies \r{d}: $\bigcup_{i \leq n} H_i = G$. 
By Lemma \ref{lem:charu}, $\pi(H_I) = \bigcap_{i \in I} \pi(H_i)$ for each $I \in \mathcal I(n)$: so we have
$\nu(H_I) = \nu(\pi^{-1}(\pi(H_I))) =  \nu(\pi^{-1}(\bigcap_{I} \pi(H_i))) = \nu(\pi^{-1}(\bigcap_{I} P_i)) = \mu(\bigcap_{I} P_i))$.
Then, by construction, for each $I \in \mathcal I(n)$, we have $\mu(\bigcap_{I} P_i)) b_\W \sim b_{H_I}$. Thus \rax{i/e} delivers $\mu(Q) = \sum_{\mathcal I(n)} (-1)^{|I|+1} \mu(\cap_I P_i)$, establishing (1). (2) follows immediately by the monotonicity of the Choquet integral and the fact that $\Pi$ is closed under finite unions as prescribed by \r{d}.  

Now, Theorem 3.15 of \cite{konig2009measure} states that a grounded set function satisfying (1) and (2) admits a modular (i.e., additive) extension, hence $\mu$ admits a modular extension to $2^\W$, which since $\mu(\emptyset) = 0$, can be normalized to a unique probability measure. In an abuse of notation, we will also call this $\mu$.

Finally, let $\mu(\pi(H) \triangle \pi(G)) = 0$. Notice this implies that $\mu(\pi(H \cup G))  = \mu(\pi(H) \cup \pi(G)) = \mu(\pi(H)) = \mu(\pi(G))$. By \rax{chq}, this means $G \im H$ and $H \im G$. Since $\im$ is derived from $\pi$, it must be that $\pi(H) = \pi(G)$. So $\<\pi,\mu\>$ meets the requirement \eqref{heucond} and therefore is an HEU representation of $\s$.

\textbf{Uniqueness.} The uniqueness of $\pi$ follows from the fact that $\im$ is derived from $\pi$ (Proposition \ref{prop:derivedfromrep}) and the fact that such an interpretation is unique (Proposition \ref{prop:drv}). The uniqueness of $\mu$, over the algebra generated by $\pi(2^\W)$ follows from the fact that $\Pi$ is $\cap$-closed and the well known fact that if two measures coincide on a $\cap$-closed family of sets then they coincide on the sigma-algebra generated by that family. 
\end{tproof}

\begin{pproof}{prop:cs}
By Proposition \ref{prop:derivedfromrep}, $\im_i$ is derived from $\<\pi_i,\mu_i\>$. for $i  \in \{1,2\}$.

\textbf{(1 implies 2)} Assume $\pi_1(H) \subseteq \pi_2(H)$ for all $H$. Let $G \im_1 H$. Then by Lemma \ref{lem:charstar}, $G \subseteq \pi_1(H)$ and so $G \subseteq \pi_2(H)$. Leaning on Lemma \ref{lem:charstar} again, $G \im_2 H$. 

\textbf{(2 implies 1)} Now assume $G \im_1 H$ implies $G \im_2 H$ for all $G,H$. By Lemma \ref{lem:charstar}, $\pi_1(H) \im_1 H$, so by assumption $\pi_1(H) \im_2 H$, and so by, you guessed it, Lemma \ref{lem:charstar}, $\pi_1(H) \subseteq \pi_2(H)$.
\end{pproof}

\newpage

\bibliographystyle{ifac.bst}
\singlespace
\bibliography{HT.bib}

\end{document}